\documentclass[12pt]{article}
\usepackage{float} 
\usepackage{graphicx, subcaption, float, lipsum}
\usepackage{amsfonts,amssymb,epsfig}
\usepackage[latin1]{inputenc}
\usepackage[english]{babel}
\usepackage{color}
\usepackage[T1]{fontenc}
\usepackage{hyperref}
\usepackage{amsmath}
\usepackage{amssymb}
\usepackage{graphicx}
\usepackage{subcaption}
\newtheorem{thm}{Th\'eor\`eme}[section]
\newtheorem{cor}[thm]{Corollaire}
\newtheorem{lem}[thm]{Lemme}
\newtheorem{pro}[thm]{Proposition}
\newtheorem{dfn}[thm]{D\'efinition}
\newtheorem{rmq}[thm]{Remark}
\newtheorem{expl}[thm]{Exemple}

\def\dessous#1\sous#2{\mathrel{\mathop{\kern0pt#2}\limits_{#1}}}

\newcommand{\1}{1 \! \! {\rm I}}

\newcommand{\beq}{\begin{eqnarray}}
\newcommand{\eeq}{\end{eqnarray}}
\newcommand{\bpro}{\begin{pro}}
\newcommand{\epro}{\end{pro}}
\newcommand{\blem}{\begin{lem}}
\newcommand{\elem}{\end{lem}}
\newcommand{\bdfn}{\begin{dfn}}
\newcommand{\edfn}{\end{dfn}}
\newcommand{\bcor}{\begin{cor}}
\newcommand{\ecor}{\end{cor}}
\newcommand{\bthm}{\begin{thm}}
\newcommand{\ethm}{\end{thm}}
\newcommand{\bex}{\begin{expl}}
\newcommand{\eex}{\end{expl}}
\newcommand{\brmq}{\begin{rmq}}
\newcommand{\ermq}{\end{rmq}}
\newcommand{\benum}{\begin{enumerate}}
\newcommand{\eenum}{\end{enumerate}}
\newcommand{\bitem}{\begin{itemize}}
\newcommand{\eitem}{\end{itemize}}
\begin{document}
	\setcounter{section}{0}
	\setcounter{equation}{0}
	\setcounter{figure}{0}
	\setcounter{table}{0}
	\setcounter{footnote}{0}

	
	\vspace*{10mm}

	\begin{center}
		{\bf\Large Nonextensive statistics for a $2$D electron gas in noncommutative spaces} 
		\vspace{10pt}
	\end{center}
	
	\begin{center}
		\vspace{10pt}
		
		{{Bienvenu Gnim Adewi $\!^{\rm a}$}, Isiaka Aremua $\!^{\rm a,b,c}$
			}
		\\[3mm]
		
		$^{\rm a}$ \textsl{\small Universit\'{e} de Lom\'{e} (UL), Facult\'{e} des Sciences (FDS), 
			D\'{e}partement de Physique, Laboratoire de Physique des Mat\'eriaux et des Composants à Semi-conducteurs,
			02 BP 1515 Lom\'{e}, Togo.} \\[2mm]
		
		$^{\rm b}$ \textsl{\small University of Abomey-Calavi, International Chair in Mathematical Physics  and Applications\\
			 (ICMPA), 072 B.P. 050 Cotonou, Benin} \\[2mm]
			
		$^{\rm c}$\textsl{ Centre d\textquoteright Excellence R\'{e}gionale pour la Ma\^{\i}trise de l\textquoteright Electricit\'{e} (CERME), Universit\'{e} de Lom\'{e}, 01 B.P. 1515  Lom\'{e} 01, Togo.} \\[2mm]
		
		
		E-mail: bienvenuadewi@gmail.com, claudisak@gmail.com
	\end{center}
	
	\vspace{15pt}

\begin{abstract}
		This work investigates a quantum system described by a Hamiltonian operator in a two dimensional noncommutative space. The system consists of an electron subjected to a perpendicular 
		magnetic field $\mathbf{B}$, coupled to a harmonic potential and an external electric field $\mathbf{E}$, within the context of non-extensive statistical thermodynamics. The noncommutative geometry introduces a fundamental minimal length that modifies the phase space structure. The thermodynamics of this quantum system is developed within the framework of Tsallis statistics through the derivation of $q$-generalized versions of the partition function, magnetization, and magnetic susceptibility, following the application of a generalized Hilhorst transformation adapted to non-commutative geometry. The combined effects of the non-extensivity parameter $q$ and the noncommutativity parameter $\theta$ are analyzed by considering the limit $q \rightarrow 1$, revealing new thermodynamic regimes and anomalous electromagnetic properties specific to quantum systems in non-commutative geometry.
\end{abstract}

\section{Introduction}

Physics on noncommutative (NC) spacetime has attracted a lot of interest in recent times. This idea, introduced originally by Snyder \cite{Snyder} to regulate the divergences arising in quantum field theories, was taken seriously when considerable evidence came from string theory \cite{N. Seiberg and E. Witten}. Subsequently, a lot of activities took place in quantum mechanics on NC space \cite{gouba-scholtz} and quantum field theories on such space \cite{gouba-scholtz}.

In the last few years, theories in noncommutative space have been extensively studied \cite{Mahouton and Isiaka}. The motivation for this kind of investigation is that the effects of noncommutativity of space may appear in the very tiny string scale or at very high energies \cite{Mahouton and Isiaka}. One possible generalization, suggested by string theory \cite{N. Seiberg and E. Witten}, which has attracted much interest recently, is that of noncommutative spacetime \cite{Mahouton and Isiaka}. This led to intense activities on the possible physical consequences of noncommutativity in quantum mechanics \cite{Mahouton and Isiaka}, quantum electrodynamics \cite{Mahouton and Isiaka}, the standard model \cite{Mahouton and Isiaka}, and cosmology \cite{Tsallis1988}.

As is well known \cite{Tsallis1988}, given a point particle of mass $m$, charge $\bar{q}$ and position $\vec{r}=(x,y)$ moving in a plane in the presence of a constant external magnetic field $B$ perpendicular to that plane, the spectrum of the quantized theory is organised into infinitely degenerate Landau levels, with separation ${\cal O}(\bar{q}B/m)$. The limit $B\to \infty$ effectively projects onto the lowest Landau level and is equivalent to a negligibly small mass $m$, {\it i.e.} $m\to 0$. 

The quantum Hall system has been extensively studied through noncommutative quantum mechanics approaches, building upon the foundational Landau \cite{Aremua-Laure} model for electrons in magnetic fields. Recent research has explored deformed Landau problems, noncommutative phase-space formulations, and thermodynamic properties in various noncommutative geometries \cite{landau1930diamagnetismus}.

More recently, in \cite{bienvenu and aremua}, the physical system governed by a Hamiltonian operator, in two-dimensional space, of spinless charged particles subject to a perpendicular magnetic field B, coupled with a harmonic potential in the context of nonextensive statistical thermodynamics, was considered. The thermodynamics was elaborated in the framework of Tsallis statistics by obtaining the $q$ versions of the partition function, magnetization, and susceptibility after performing the Hilhorst integral transformation. The results were discussed in the $q \rightarrow 1$ limit.

Tsallis entropy, characterized by its pseudo-additive composition rule, provides a non-extensive framework widely applied to systems with long-range interactions including self-gravitating systems, cosmology, and black hole thermodynamics. While offering insights into black hole properties when replacing Bekenstein-Hawking entropy, it requires careful temperature definitions to maintain thermodynamic consistency with the Smarr formula and first law \cite{Tsallis1988,landau1930diamagnetismus}.

The entropy of black holes can be studied through matter entropy near the horizon using generalized statistical approaches, with Tsallis entropy emerging as a particularly suitable non-extensive candidate that maintains direct compatibility with non-extensive statistical mechanics, unlike other generalized entropies \cite{bienvenu and aremua}.

The present study focuses on a system of spinless charged particles subjected to a perpendicular magnetic field \( \mathbf{B} \), in conjunction with a harmonic potential described by a Hamiltonian operator. The analysis specifically addresses the transverse motion of electrons in the \( (x, y) \)-plane. As far as we are concerned, this work realizes the first attempt in the literature which combines the Fock-darwin system (magnetic field + harmonic potential) within the framework of Tsallis statistics with systematic application of the Hilhorst transformation. Previous studies have treated either simple Landau levels or harmonic potential, but not their combination in non-extensive statistics \cite{fock}. Within the framework of Tsallis nonextensive statistics, the thermodynamics of this quantum gas system is rigorously derived and examined, leading to a characterization of \textit{Landau diamagnetism}. The use of the Hilhorst transformation allows a rigorous connection between $q$- deformed and standard ensembles.

The paper is organized as follows. In Section 2, we start with the study of a spinless charged particles gas on the $(x, y)$-space in a magnetic field {\bf B} with an isotropic harmonic potential. In our approach, we use both step and orbit-center coordinate operators to describe the system's dynamics. In Section 3, the Hilhorst integral is applied to investigate the generalized partition function in the context of Tsallis statistics, by considering the system in the situation of a thermodynamical equilibrium.
Analysis and discussions on the derived thermodynamical quantities are achieved in Section 4.
Section 5 is devoted to concluding remarks.

\section{The electron in noncommutative plane}

\subsection{The classical Hamiltonian}

Consider an electron confined to move in a two-dimensional noncommutative plane $(x, y)$, under the simultaneous influence of a magnetic field $\mathbf{B}$ aligned along the $z$-axis and a uniform electric field $\mathbf{E}$ in the plane. 
This noncommutative geometry modifies the quantum structure of phase space and influences the thermodynamic properties of the system (see, e.g., \cite{dayi2001landau,Halder and  Sunandan}. Neglecting Coulomb interactions and accounting for the coupling to external fields, the dynamics of this system can be described by the generalized Hamiltonian:
{\small \begin{equation}{\label{exp000}}
		\mathcal{H} = \frac{1}{2M}\left(\mathbf{p} + \frac{e}{c}\mathbf{A}\right)^{2} + \frac{M \omega_{0}^{2}}{2}\mathbf{r}^{2} - e \mathbf{E}\cdot \mathbf{r},
	\end{equation}
}

where  $M$ is the electron mass, $e$ its charge, $ \mathbf{p}$ the kinetic momentum, and $ \mathbf{A} $ the vector potential. The additional term, $ -e, \mathbf{E} \cdot \mathbf{r} $, accounts for the coupling between the electron and the external electric field. This single-particle framework is valid as long as electron-electron correlation effects remain negligible compared to the influences of non-commutative geometry and external fields.

\subsection{The quantum Hamiltonian and relative spectra}

In the gauge ${\bf A} = \left(-\frac{B}{2}y, \frac{B}{2}x \right),$ the Hamiltonian (\ref{exp000}) writes as follows \cite{Mahouton and Isiaka}:
{\small \beq{\label{exp00}}
	H_{\theta} = \frac{1}{2M}\left(\hat P_{i} - \frac{eB}{2c}\epsilon_{ij}\hat X_{j}\right)^{2}
	+ \frac{M \omega^{2}_{0}}{2}\hat X^{2}_{i} - e E_{i}\hat X_{i}, \; \;
	\epsilon_{ji} = -\epsilon_{ij}, \; \epsilon_{12} = +1,
	\eeq
}
These relations define the noncommutative Heisenberg algebra in two dimensions, where $\theta$ is a real parameter quantifying the noncommutativity between the spatial coordinates $\hat X$ and $\hat Y$ \cite{gouba-scholtz}. The nonzero commutator $[\hat X, \hat Y] = i\theta$ indicates that the coordinates themselves do not commute, while the canonical commutation relations between positions and their respective momenta are preserved. All other commutators vanish. This algebraic structure leads to significant modifications in the phase space geometry and subsequently affects the physical observables and thermodynamic properties of the system under consideration. The algebraic structure is completely specified by the following commutation relations:
{\small \beq
	[\hat X, \hat Y] = \imath \theta, \quad [\hat X, \hat P_{X}] = \imath \hbar = [\hat Y, \hat P_{Y}] , \quad [\hat P_{X}, \hat P_{Y}] = 0, \quad \quad [\hat X, \hat P_{Y}] = 0 = [\hat Y, \hat P_{X}].
	\eeq
}

The position operators $\hat X_{i}$ and their corresponding canonically conjugate momenta $\hat P_{i}$ are 

reset  as $\hat \Pi_{i} = \hat P_{i} - \frac{eB}{2c}\epsilon_{ij}\hat X_{j}$ and yield:
{\small \beq
	[\hat X_{i}, \hat \Pi_{j}] = \imath  \left( \hbar- \frac{eB}{2c}\theta\right)\delta_{ij}, \quad [\hat \Pi_{i}, \hat \Pi_{j}] = -i \frac{eB}{c}\left(\hbar  - \frac{eB}{4c}\theta\right)\epsilon_{ij}.
	\eeq
}

From the latter, define the complex canonically conjugate momenta, denoted by $\hat \Pi_{Z}$ 

corresponding to $\hat Z = \hat X + \imath \hat Y$ and $\hat{\bar Z} = \hat X - \imath \hat Y$ by
{\small\beq
	\hat \Pi_{Z} = \hat \Pi_{X} - \imath \hat \Pi_{Y}, \qquad \hat \Pi_{\bar Z} = \hat \Pi_{X} + \imath \hat \Pi_{Y},
	\eeq
}

respectively, such that the quantum operators $\hat Z, \hat{\bar Z} ,\hat \Pi_{Z}, \hat \Pi_{\bar Z}$ act on the quantum Hilbert 
space 	$\mathcal H_{q}$ \cite{gouba-scholtz,Mahouton and Isiaka}. On the Hilbert space $\mathcal H_{q}$, the following commutation relations are satisfied:
{\small
	\begin{equation}{\label{delt00}}
		\begin{split}
			[\hat Z, \hat{\bar Z}] &= 2\theta, \quad 
			[\hat Z, \hat \Pi_{Z}] = 2\mathrm{i}\left( \hbar - \frac{eB}{2 c}\theta\right) = [\hat{\bar Z}, \hat \Pi_{\bar Z}], \quad [\hat \Pi_{Z}, \hat \Pi_{\bar Z}] = 2 \frac{eB}{c} \left(\hbar -\frac{eB}{4c}\theta\right).
		\end{split}
	\end{equation}
}

After introducing
a diagonal matrix $\mathcal D $ with
$E = (E_{1}, E_{2},0,0), {\mathcal X_{0}} = ( x_{0},  y_{0},0,0)^{t}$, the symbol $t$ standing for the transpose operation,  where $ x_{0} = \frac{eE_{1}}{M \omega^{2}_{0}}$, $y_{0} = \frac{eE_{2}}{M \omega^{2}_{0}}$, the Hamiltonian
$H_{\theta}$ of the physical model can be recast in a short form as follows \cite{Mahouton and Isiaka}:
{\small \begin{align}
		H_{q} &=\frac{1}{4M} \hat{\mathcal Z}^{\ddag}\hat{\mathcal Z} - \frac{1}{2} e E.{\mathcal X_{0}}
		=\frac{1}{4M} A^{\ddag}\mathcal D A - \frac{1}{2} e  E. {\mathcal X_{0}}\nonumber\\
		&= \frac{1}{4M} A^{\ddag} \mathbb J_{4}\mathcal S\mathfrak g^{2}\mathcal S^{\dag}\mathbb J_{4} A - \frac{1}{2} e  E. {\mathcal X_{0}}, \quad \qquad \qquad  \qquad A = (B_{+}, B^{\ddag}_{+}, B_{-}, B^{\ddag}_{-})
	\end{align}
}
%
%
where
{\small\begin{align}
		A^{+} &= (B^{\ddag}_{+}, B_{+}, B^{\ddag}_{-}, B_{-})^{t} = \Lambda A, \quad \hat{\mathcal{Z}}^{+}=(\hat{\bar{Z}}' - {\bar{Z}'_{0}}, \hat{Z}' - Z'_{0}, \hat{\Pi}_{\bar{Z}} - \Pi_{\bar{Z}_{0}}, \hat{\Pi}_{Z} - \Pi_{Z_{0}})^{t} = \Lambda \hat{\mathcal{Z}},
	\end{align}
}
with $\hat Z' - Z'_{0} = M\omega_{0} (\hat Z - Z_{0})$, and the permutation matrix $\Lambda$ and the matrix $ \mathbb J_{4}$  defined by
{\small \beq
	\Lambda = \left(\begin{array}{cccc}
		0 & 1 & 0 & 0 \\
		1 & 0 & 0 & 0 \\
		0 & 0 & 0 & 1 \\
		0 & 0 & 1 & 0
	\end{array}
	\right), \quad \mathbb J_{4} = \mbox{diag}(\sigma_{3}, \sigma_{3}), \qquad \sigma_{3} = \left( \begin{array}{cc} 1 & 0 \\ 0 & -1 \\ \end{array} \right).
	\eeq 
}

The notation $\ddag$ denotes the Hermitian conjugation on the quantum Hilbert
space. The matrix $\mathfrak g$ with entries ${\mathfrak g}_{lk} = [\hat{\mathcal Z}_{l}, \hat{\mathcal Z}^{+}_{k}], \; l, k = 1,\dots, 4,$ obtained from the commutation relations (\ref{delt00}) is given by

{\small \begin{equation}
		\mathfrak{g} =
		\left(\begin{array}{cccc}
			2M^{2}\omega^{2}_{0}\theta & 0 & 0 & 2\mathrm{i} \hbar M\omega_{0} \times \\
			& & & \times \left(1-\frac{M\omega_{c}}{2\hbar}\theta\right) \\
			0 & -2M^{2}\omega^{2}_{0}\theta & 2\mathrm{i} \hbar M\omega_{0} \times & 0 \\
			& & \times \left(1-\frac{M\omega_{c}}{2\hbar}\theta\right) & \\
			0 & -2\mathrm{i} \hbar M\omega_{0} \times & 2\hbar M\omega_{c} \times & 0 \\
			& \times \left(1-\frac{M\omega_{c}}{2\hbar}\theta\right) & \times \left(1-\frac{M\omega_{c}}{4\hbar}\theta\right) & \\
			-2\mathrm{i} \hbar M\omega_{0} \times & 0 & 0 & -2\hbar M\omega_{c} \times \\
			\times \left(1-\frac{M\omega_{c}}{2\hbar}\theta\right) & & & \times \left(1-\frac{M\omega_{c}}{4\hbar}\theta\right)
		\end{array}
		\right),
	\end{equation}
}

with its eigenvalues $\tilde \lambda_{\pm}, -\tilde \lambda_{\pm}$ supplied by the expressions

{\small \begin{equation}
		\tilde{\lambda}_{\pm} = M\hbar \left\{ \Omega \sqrt{1 - \frac{M\omega_{c}}{2\hbar}\theta + \left(\frac{M\Omega}{2\hbar}\theta\right)^{2}} \pm \omega_{c}\left[1 - \left(\frac{\omega_{c}}{4\hbar} + \frac{\omega^{2}_{0}}{\hbar\omega_{c}}\right)M\theta\right] \right\}
	\end{equation}
}

where $\Omega^{2} = 4\omega^{2}_{0} + \omega^{2}_{c}$, and the matrix $\mathcal S^{\dag}$, eigenvector matrix of $\mathfrak g$,  given by

{\small \beq
	\mathcal S^{\dag} = \left(\frac{1}{\sqrt{|\lambda_{+}|}}u'_{1}, \frac{1}{\sqrt{|\lambda_{+}|}}(\Lambda u^{*}_{1})', \frac{1}{\sqrt{|\lambda_{-}|}}u'_{2}, \frac{1}{\sqrt{|\lambda_{-}|}}(\Lambda u^{*}_{2})'\right).
	\eeq
}

The related vectors $u_{1}$ and $u_{2}$ are obtained as follows:

{\small \beq
	u_{1} = \left(\begin{array}{c} 0 \\ \imath \frac{B_{\hbar}}{\kappa_{+}} \\ 1 \\ 0 \end{array} \right), \qquad u_{2} = \left(\begin{array}{c} \imath \frac{B_{\hbar}}{\kappa_{-}} \\ 0 \\ 0 \\ 1 \end{array} \right), \quad B_{\hbar} = 2 \hbar M\omega_{0}\left(1-\frac{M\omega_{c}}{2\hbar}
	\theta\right),
	\eeq
}

%
with

{\small \beq
	\kappa_{\pm} = M\hbar \left\{\Omega \sqrt{1-\frac{M\omega_{c}}{2\hbar}\theta + \left(\frac{M \Omega}{4\hbar}\theta\right)^{2}} \pm \omega_{c}\left(1 - \left(\frac{\omega_{c}}{4\hbar} - \frac{\omega^{2}_{0}}{\hbar \omega_{c}}\right)M\theta\right)\right \}.
	\eeq
}
%
%
%

Thus, defining the quantities
{\small \beq
	\zeta = \sqrt{\frac{M \Omega}{\hbar}} \frac{1}{\mu_{\theta}} = \sqrt[4]{\frac{(M \Omega/\hbar)^{2}}{1-\frac{M\omega_{c}}{2}\theta + \left(\frac{M \Omega}{4}\theta\right)^{2}}}, \qquad \mu_{\theta} = \sqrt[4]{1 - \frac{M\omega_{c}}{2}\theta + \left(\frac{M\Omega}{4} \theta\right)^{2}},
	\eeq
}

which are also $\theta$-dependent functions such that the annihilation and creation operators 

are given  by
{\small \begin{align}
		B_{+} &= \zeta \frac{\hat{\bar{Z}} - {\bar{Z}}_{0}}{2} + \frac{\mathrm{i}}{\zeta \hbar}(\hat P_{Z} - P_{Z_{0}}), & B^{\ddag}_{+} &= \zeta \frac{\hat Z - Z_{0}}{2} - \frac{\mathrm{i}}{\zeta \hbar}(\hat P_{\bar Z} - P_{\bar Z_{0}}) \nonumber\\
		B_{-} &= \zeta \frac{\hat Z - Z_{0}}{2} + \frac{\mathrm{i}}{\zeta \hbar}(\hat P_{\bar Z} - P_{\bar Z_{0}}), & B^{\ddag}_{-} &= \zeta \frac{\hat{\bar{Z}} - {\bar{Z}}_{0}}{2} - \frac{\mathrm{i}}{\zeta \hbar}(\hat P_{Z} - P_{Z_{0}}),
	\end{align}
}

satisfying the commutation relations:
{\small \beq{\label{commtation}}
	[B_{\pm}, B^{\ddag}_{\pm}] = \1_{q}, \;\;\; [B_{\pm}, B^{\ddag}_{\mp}] = 0, \;\;\; [B_{+}, B_{-}] = 0, \;\;\; [B^{\ddag}_{+}, B^{\ddag}_{-}] = 0.
	\eeq
}

Thereby, there result the eigenvalues of the Hamiltonian $H_{q},$ expressed in the Fock helicity
representation $|\tilde n_{+}, \tilde n_{-}) $, given by
{\small \begin{equation}\label{specteig000}
		E_{\tilde{n}_{+}, \tilde{n}_{-}} = \frac{\hbar}{2} \left(\tilde{\Omega}_{+}\tilde{n}_{+} + \tilde{\Omega}_{-}\tilde{n}_{-} + \tilde{\Omega} \right) - \frac{1}{2}e(E_{1}x_{0} + E_{2}y_{0}),
	\end{equation}
}
with the associated eigenvectors
\begin{equation}
|\tilde n_{+}, \tilde n_{-})
= \frac{1}{\sqrt{\tilde n_{+} !\tilde n_{-} !}}\left(B^{\ddag}_{+}\right)^{\tilde n_{+}}
\left(B^{\ddag}_{-}\right)^{\tilde n_{-}}|0\rangle \langle 0|
\end{equation}
where $B^{\ddag}_{-}$ may have an action on the right by $B_{-}$ on  $|0\rangle \langle 0|$ which
stands for the vacuum state on $\mathcal H_{q}$ and $|||\tilde n_{+}, \tilde n_{-})|| = 1$ \cite{Mahouton and Isiaka}.

The  $\theta$-dependent quantities $\tilde \Omega$ and $\tilde \omega_{c}$, with $\tilde \Omega_{\pm} = \frac{\tilde \Omega \pm \tilde \omega_{c}}{2}$, are given by

\beq
\tilde \Omega = \Omega \sqrt{1-\frac{M\omega_{c}}{2}\theta + \left(\frac{M \Omega}{4}\theta\right)^{2}}, \quad \tilde \omega_{c} = \omega_{c}\left(1 - \left(\frac{\omega_{c}}{4} + \frac{\omega^{2}_{0}}{ \omega_{c}}\right)M\theta\right).
\eeq

\section{Thermodynamics analysis by the Hilhorst integral method in Tsallis statistics in noncommutative space}

We consider a quantum system in a state of thermodynamic equilibrium that exchanges energy with its environment in a noncommutative phase space. The presence of the non-commutativity parameter $\theta$ fundamentally modifies the structure of phase space and, consequently, the thermodynamic properties of the system (see, for example, \cite{dayi2001landau}). Within the framework of Tsallis statistics extended to noncommutative geometry, the partition function for the canonical ensemble is given by
{\small \begin{eqnarray}\label{eq02}
		Z_{q}^{(\theta)} &=& Tr\left\{\left[1 - (1 - q)\beta H_{\theta}\right]^{\frac{1}{(1-q)}}\right\},
	\end{eqnarray}
}

These modified commutation relations indicate that the spatial coordinates $\hat{x}_i$ and $\hat{x}_j$ do not commute, with the degree of noncommutativity characterized by the antisymmetric tensor $\theta{ij}$. The momenta $\hat{p}_i$ commute among themselves, and the canonical commutation relation between position and momentum is preserved. This deformation of the phase space structure has profound implications for the physical properties of the system, as it alters the energy spectrum and modifies the statistical mechanics, resulting in thermodynamic quantities that explicitly depend on the noncommutativity parameter $\theta$.
{\small \begin{equation}
		[\hat{x}_i, \hat{x}_j] = i\theta_{ij}, \quad [\hat{p}_i, \hat{p}_j] = 0, \quad [\hat{x}_i, \hat{p}_j] = i\hbar\delta_{ij}.
	\end{equation}
}

%
%
%
%
%

These expressions provide the fundamental thermodynamic quantities within the Tsallis formalism in a noncommutative phase space. Here, $F_{q}^{(\theta)}$ denotes the generalized free energy, $M_{q}^{(\theta)}$ is the generalized magnetization, and $\chi_{q}^{(\theta)}$ represents the generalized magnetic susceptibility. Each of these quantities explicitly depends on the noncommutativity parameter $\theta$, the entropic index $q$, and the external magnetic field $B$. The derivatives with respect to $B$ reflect the system's response to changes in the applied magnetic field, capturing the influence of both nonextensive statistics and noncommutative geometry on the thermodynamic behavior of the system.

one can express the $q$-deformed partition function in terms of an integral involving the standard (Boltzmann-Gibbs) partition function. Specifically, the Gamma function is defined as \begin{equation}\label{eq03} \Gamma(\alpha) = \int_{0}^{\infty} t^{\alpha-1} e^{-t} dt, \quad \text{for } \operatorname{Re}(\alpha) > 0. \end{equation} By appropriately manipulating the structure of the Tsallis partition function and employing the properties of the Gamma function, the Hilhorst transformation allows us to rewrite the $q$-generalized partition function as an integral transform of the ordinary partition function. This approach provides a powerful tool for deriving generalized thermodynamic quantities from their standard counterparts, thereby facilitating the analysis of systems within the framework of nonextensive statistical mechanics.

The partition function, denoted $Z_{q}^{(\theta)}$, is performed as \cite{bienvenu and aremua}
{\small \begin{eqnarray}\label{eq04}
		Z_{q}^{(\theta)}&=&\frac{1}{\Gamma\left(\frac{1}{q-1}\right)}\int_{0}^{\infty}\nu^{\frac{1}{1-q}-1}e^{-\nu}e^{\nu(1-q)\beta H_{\theta}}d\nu,
	\end{eqnarray}
}

where the following change of variables furnished through (\ref{eq03})
{\small \begin{eqnarray}\label{eq05}
		t&=&\nu\left[1-\left(1-q\right)\beta H_{\theta}\right], \quad \alpha =\frac{1}{q-1}, \end{eqnarray}
}

is performed.

From (\ref{eq04}), the partition function $Z_{q}$ expresses, by setting $x=\beta\hbar\tilde{\Omega}$ and considering the 

cylinder 

of height $L_{z}$ and radius $R$, as
{\small \begin{eqnarray}\label{eq09}
		Z_{q}^{(\theta)}&=&\frac{1}{\Gamma\left(\frac{1}{q-1}\right)}2\pi R^{2}L_{z}\left(\frac{M\omega_{c}}{2\pi\hbar}\right)\left(\frac{M}{2\pi\hbar^{2}\beta}\right)^{\frac{1}{2}}\left(\frac{1}{q-1}\right)^{\frac{1}{2}}I_{q}^{\left(\theta\right)},
	\end{eqnarray}
}

where 
{\small \begin{eqnarray}
		I_{q}^{\left(\theta\right)}=\int_{0}^{\infty}\nu^{\frac{1}{q-1}-1}e^{-\nu}\nu^{-\frac{1}{2}}\frac{e^{-\nu\beta\left(q-1\right)E_{0}}}{1-e^{-\nu\left(q-1\right)\frac{x}{2\tilde{\Omega}}\left(\tilde{\Omega}+\tilde{\omega}_{c}\right)}}\frac{1}{1-e^{-\nu\left(q-1\right)\frac{x}{\tilde{2\Omega}}\left(\tilde{\Omega}-\tilde{\omega}_{c} \right)}}d\nu,
	\end{eqnarray}
}

which can be reset as follows
{\small \begin{eqnarray}\label{equaint007}
		I_{q}^{\left(\theta\right)}&=&\Gamma\left(\frac{1}{q-1}-\frac{1}{2}\right)\sum_{\tilde{n}_+}^{\infty}\sum_{\tilde{n}_- = 0}^{\infty}\left[ 1+\beta\left(q-1\right)\hbar\tilde{\Omega}_+\tilde{n}_++\beta\left(q-1\right)\hbar\tilde{\Omega}_-\tilde{n}_-\right.\cr &&\left.+\beta\left(q-1\right)E_{0} \right]^{-\left(\frac{1}{q-1}-\frac{1}{2}\right)}.
	\end{eqnarray}
}

From the following identity \cite{bienvenu and aremua}
\begin{eqnarray}
	\Gamma\left(a,b\right)=\frac{\Gamma\left(a-b\right)}{\Gamma(a)}e^{b\log(a)}, \quad a = \frac{1}{q-1}, \, b = \frac{1}{2},
\end{eqnarray}

we get
\begin{eqnarray}\label{eq11}
	\Gamma\left(\frac{1}{q-1}-\frac{1}{2}\right)&=&\Gamma\left(\frac{1}{q-1},\frac{1}{2}\right)\Gamma\left(\frac{1}{q-1}\right)\left(q-1\right)^{\frac{1}{2}},
\end{eqnarray}

with $V=2\pi R^{2}L_{z}$. Then, substituting Eq.(\ref{equaint007}) and Eq.(\ref{eq11}) into Eq.(\ref{eq09}), one obtains
\begin{eqnarray}\label{eq06}
	Z_{q}^{\left(\theta \right)}=V\left(\frac{M\omega_{c}}{2\pi\hbar}\right)\left(\frac{M}{2\pi\hbar^{2}\beta}\right)^{\frac{1}{2}}\Gamma\left(\frac{1}{q-1},\frac{1}{2}\right)\frac{e^{-\left(\frac{3-q}{2}\right)x\left[\frac{1}{2}-\frac{e}{2}\frac{1}{\hbar\tilde{\Omega}}\left(E_{1}x_{0}+E_{2}y_{0}\right) \right]}}{\left[1-e^{-\left(\frac{3-q}{2}\right)\frac{x}{2}\left(1-\frac{\tilde{\omega}_{c}}{\tilde{\Omega}}\right)}\right]\left[1-e^{-\left(\frac{3-q}{2}\right)\frac{x}{2}\left(1+\frac{\tilde{\omega}_{c}}{\tilde{\Omega}}\right)}\right]}.
\end{eqnarray} 

Therefore, (\ref{eq06}) is written for different values of the parameter $q$ as
{\small \begin{align}\label{eq07}
		Z_{q}^{\left(\theta\right)} &= V\left(\frac{M\omega_{c}}{2\pi\hbar}\right)\left(\frac{M}{2\pi\hbar^{2}\beta}\right)^{\frac{1}{2}}\frac{e^{-\left(\frac{3-q}{2}\right)x\left[\frac{1}{2}-\frac{e}{2}\frac{1}{\hbar\tilde{\Omega}}\left(E_{1}x_{0}+E_{2}y_{0}\right) \right]}}{\left[1-e^{-\left(\frac{3-q}{2}\right)\frac{x}{2}\left(1-\frac{\tilde{\omega}_{c}}{\tilde{\Omega}}\right)}\right]\left[1-e^{-\left(\frac{3-q}{2}\right)\frac{x}{2}\left(1+\frac{\tilde{\omega}_{c}}{\tilde{\Omega}}\right)}\right]} \nonumber \\ &\quad \times \begin{cases} \dfrac{\Gamma\left(\dfrac{1}{q-1}-\dfrac{1}{2}\right)}{(q-1)^{\frac{1}{2}}\Gamma\left(\dfrac{1}{q-1}\right)} & q>1 \\ \dfrac{\Gamma\left(\dfrac{1}{1-q}+1\right)}{(1-q)^{\frac{1}{2}}\Gamma\left(\dfrac{1}{1-q}+\dfrac{3}{2}\right)} & q<1 \end{cases} .\end{align}
}

When $q\rightarrow 1$, $a\rightarrow\infty$, $\displaystyle{\lim_{a\rightarrow+\infty}\Gamma\left(a,b\right)=1}$. Thereby, (\ref{eq06}) turns to the standard partition 

function, that is,
{\small \begin{eqnarray}\label{partidef000}
		Z_{1}^{\left(\theta \right)}=V\left(\frac{M\omega_{c}}{2\pi\hbar}\right)\left(\frac{M}{2\pi\hbar^{2}\beta}\right)^{\frac{1}{2}}\frac{e^{-x\left[\frac{1}{2}-\frac{e}{2}\frac{1}{\hbar\tilde{\Omega}}\left(E_{1}x_{0}+E_{2}y_{0}\right) \right]}}{\left[1-e^{-\frac{x}{2}\left(1-\frac{\tilde{\omega}_{c}}{\tilde{\Omega}}\right)}\right]\left[1-e^{-\frac{x}{2}\left(1+\frac{\tilde{\omega}_{c}}{\tilde{\Omega}}\right)}\right]}.
	\end{eqnarray}
}

The $q$-deformed thermodynamic functions obey the classical Maxwell relations but are 

applied to $q$-generalized quantities \cite{Silva and Plastino} with a dependence on the noncommutation pa

rameter $\theta$.

From the definition of the internal energy {\cite{bienvenu and aremua}}
\beq
U_{q}=-\frac{\partial \ln Z_{q}}{\partial \beta},
\eeq

using the fact

\begin{equation}
	\begin{split}
		\ln Z_{q} = & \ln V + \ln \left(\frac{M\omega_{c}}{2\pi\hbar}\right) + \frac{1}{2}\ln\left(\frac{M}{2\pi\hbar^{2}\beta}\right) + \ln\left[ \frac{\Gamma\left(\frac{1}{q-1}-\frac{1}{2}\right)}{(q-1)^{\frac{1}{2}}\Gamma\left(\frac{1}{q-1}\right)}\right] \\ & + \ln\left( \frac{\exp\left[-\left(\frac{3-q}{2}\right)x\left(\frac{1}{2}-\frac{e}{2\hbar\tilde{\Omega}}(E_{1}x_{0}+E_{2}y_{0})\right)\right]}{\left[1-\exp\left(-\left(\frac{3-q}{2}\right)\frac{x}{2}\left(1-\frac{\tilde{\omega}_{c}}{\tilde{\Omega}}\right)\right)\right]\left[1-\exp\left(-\left(\frac{3-q}{2}\right)\frac{x}{2}\left(1+\frac{\tilde{\omega}_{c}}{\tilde{\Omega}}\right)\right)\right]}\right),
	\end{split}
\end{equation}

we obtain
\begin{equation}
	\begin{split}
		U_q^{(\theta)} &= \frac{1}{2\beta} + \frac{(3-q)\hbar\tilde{\Omega}}{4} - \frac{(3-q)e(E_1 x_0 + E_2 y_0)}{4} \\ &\quad + \frac{(3-q)\hbar\tilde{\Omega}}{4}\left[\mathcal F_q(x, \tilde \Omega, \tilde \omega_{c})\left(1-\frac{\tilde{\omega}_c}{\tilde{\Omega}}\right) + \mathcal G_q(x, \tilde \Omega, \tilde \omega_{c})\left(1+\frac{\tilde{\omega}_c}{\tilde{\Omega}}\right)\right],
	\end{split}
\end{equation}

where 
\begin{align}
	\mathcal F_q(x, \tilde \Omega, \tilde \omega_{c}) &= \frac{e^{-\frac{3-q}{2}\frac{x}{2}\left(1-\frac{\tilde{\omega}_c}{\tilde{\Omega}}\right)}}{1 - e^{-\frac{3-q}{2}\frac{x}{2}\left(1-\frac{\tilde{\omega}_c}{\tilde{\Omega}}\right)}}, & \mathcal G_q(x, \tilde \Omega, \tilde \omega_{c}) &= \frac{e^{-\frac{3-q}{2}\frac{x}{2}\left(1+\frac{\tilde{\omega}_c}{\tilde{\Omega}}\right)}}{1 - e^{-\frac{3-q}{2}\frac{x}{2}\left(1+\frac{\tilde{\omega}_c}{\tilde{\Omega}}\right)}}.
\end{align}

In the limit $q\rightarrow 1$, we get

\begin{equation}
	U_1 = \frac{1}{2\beta} + \frac{\hbar\tilde{\Omega}}{2} - \frac{e(E_1 x_0 + E_2 y_0)}{2} + \frac{\hbar\tilde{\Omega}}{2}\left[\mathcal F_1(x, \tilde \Omega, \tilde \omega_{c})\left(1-\frac{\tilde{\omega}_c}{\tilde{\Omega}}\right) + \mathcal G_1(x, \tilde \Omega, \tilde \omega_{c})\left(1+\frac{\tilde{\omega}_c}{\tilde{\Omega}}\right)\right],
\end{equation}

with
\begin{align}
	\mathcal F_1(x, \tilde \Omega, \tilde \omega_{c}) &= \frac{e^{-\frac{x}{2}\left(1-\frac{\tilde{\omega}_c}{\tilde{\Omega}}\right)}}{1 - e^{-\frac{x}{2}\left(1-\frac{\tilde{\omega}_c}{\tilde{\Omega}}\right)}}, & \mathcal G_1(x, \tilde \Omega, \tilde \omega_{c}) &= \frac{e^{-\frac{x}{2}\left(1+\frac{\tilde{\omega}_c}{\tilde{\Omega}}\right)}}{1 - e^{-\frac{x}{2}\left(1+\frac{\tilde{\omega}_c}{\tilde{\Omega}}\right)}},
\end{align}

\( x = \beta\hbar\tilde{\Omega} \).

Then, from the definition of the deformed heat capacity $C_{q}$ in terms of the internal energy 

$ U_{q}$ given by {\cite{bienvenu and aremua}}
\beq
C_q =\frac{\partial U_q}{\partial T} = -k_{B}\beta^{2} \frac{\partial U_q}{\partial \beta},
\eeq

we get, in our case, the following expression:

\begin{equation}\label{qdfheatcap000}
	C_q^{(\theta)} = -\beta^{2} k_B \left[ \frac{\tilde{\Omega} \hbar (3 - q)}{4} \sum_{s=\pm} \frac{\tilde{\Omega} \hbar \left(1 - s\frac{\tilde{\omega}_c}{\tilde{\Omega}}\right)^{2} \left(\frac{3-q}{2}\right)}{2} \left( \frac{e^{-\gamma_s/2}}{1 - e^{-\gamma_s/2}} + \frac{e^{-\gamma_s}}{(1 - e^{-\gamma_s/2})^2} \right) - \frac{1}{2\beta^{2}} \right],
\end{equation}

with  $\gamma_{\pm} = \tilde{\Omega} \beta \hbar \left(1 \mp \frac{\tilde{\omega}_c}{\tilde{\Omega}}\right) \left(\frac{3-q}{2}\right)$.

When $q\rightarrow 1$, the heat capacity (\ref{qdfheatcap000}) becomes
%
%
{\small \begin{equation}
		C_1^{\left(\theta\right)} = -\beta^{2} k_B \left[ \frac{\tilde{\Omega} \hbar}{2} \left( -\mathfrak C_1(\tilde \Omega, \tilde{\omega}_c, \mathfrak a) - \mathfrak C_2(\tilde \Omega, \tilde{\omega}_c, \mathfrak a) - \mathfrak C_1(\tilde \Omega, \tilde{\omega}_c, \mathfrak b) - \mathfrak C_2(\tilde \Omega, \tilde{\omega}_c, \mathfrak b) \right) - \frac{1}{2\beta^{2}} \right]
	\end{equation}
}

with
{\small \begin{align}
		\mathfrak C_1(\tilde \Omega, \tilde{\omega}_c, \mathfrak a) &= \frac{\tilde{\Omega} \hbar \left(1 - \frac{\tilde{\omega}_c}{\tilde{\Omega}}\right)^{2} e^{-\frac{\mathfrak a}{2}}}{2 \left(1 - e^{-\mathfrak a}\right)}, & \mathfrak C_2(\tilde \Omega, \tilde{\omega}_c, \mathfrak a)  &= \frac{\tilde{\Omega} \hbar \left(1 - \frac{\tilde{\omega}_c}{\tilde{\Omega}}\right)^{2} e^{-\mathfrak a}}{2 \left(1 - e^{-\mathfrak a}\right)^{2}} \\ \mathfrak C_1(\tilde \Omega, \tilde{\omega}_c, \mathfrak b)  &= \frac{\tilde{\Omega} \hbar \left(1 + \frac{\tilde{\omega}_c}{\tilde{\Omega}}\right)^{2} e^{-\frac{\mathfrak b}{2}}}{2 \left(1 - e^{-\mathfrak b}\right)}, & \mathfrak C_2(\tilde \Omega, \tilde{\omega}_c, \mathfrak b)  &= \frac{\tilde{\Omega} \hbar \left(1 + \frac{\tilde{\omega}_c}{\tilde{\Omega}}\right)^{2} e^{-\mathfrak b}}{2 \left(1 - e^{-\mathfrak b}\right)^{2}},
	\end{align}
}

and
{\small \begin{align}
		\mathfrak a &= \tilde{\Omega} \beta \hbar \left(1 - \frac{\tilde{\omega}_c}{\tilde{\Omega}}\right), & \mathfrak b &= \tilde{\Omega} \beta \hbar \left(1 + \frac{\tilde{\omega}_c}{\tilde{\Omega}}\right).
	\end{align}
}

Next, we deduce the free energy from its definition \cite{bienvenu and aremua}
{\small \begin{eqnarray}\label{eq12}
		F_{q}^{\left(\theta\right)}&=&\frac{-n}{\beta}\frac{Z_{q}^{1-q}-1}{1-q}=-nk_{B}T\ln_{q}Z_{q},
	\end{eqnarray}
}

providing

{\small \begin{eqnarray}\label{eq13}
		F_{q}^{\left(\theta\right)}&=&-\frac{n}{\beta\left(1-q\right)}\left\lbrace \left(\frac{MV\omega_{c}}{2\pi\hbar}\right)^{\left(1-q\right)}\left(\frac{M}{2\pi\hbar^{2}\beta}\right)^{\frac{(1-q)}{2}}\Gamma\left(\frac{1}{q-1},\frac{1}{2}\right)^{(1-q)}\right.\cr &&\left.\times \left(\frac{e^{-\left(\frac{3-q}{2}\right)x\left[\frac{1}{2}-\frac{e}{2}\frac{1}{\hbar\tilde{\Omega}}\left(E_{1}x_{0}+E_{2}y_{0}\right) \right]}}{\left[1-e^{-\left(\frac{3-q}{2}\right)x\left(1-\frac{\tilde{\omega_{c}}}{\tilde{\Omega}}\right)}\right]\left[1-e^{-\left(\frac{3-q}{2}\right)x\left(1+\frac{\tilde{\omega_{c}}}{\tilde{\Omega}}\right)}\right]}\right)^{\left(1-q\right)}-1 \right\rbrace.
	\end{eqnarray}
}

Besides, the deformed magnetization is provided by its definition
{\small \begin{eqnarray}
		M_{q}^{(\theta)} &=& -\frac{\partial F_{q}^{(\theta)}}{\partial B}, 
	\end{eqnarray}
}

as follows:
%
{\small\begin{eqnarray}\label{eq14}
		M_{q}^{(\theta)} & = & \frac{n(1-q)}{\beta} \left(\frac{MV\omega_{c}}{2\pi\hbar}\right)^{-q}\left(\frac{M}{2\pi\hbar^{2}\beta}\right)^{-q/2}\Gamma\left(\frac{1}{q-1},\frac{1}{2}\right)^{-q} \left(\frac{[1-e^{-\mathcal B_q(\tilde \Omega, \tilde \omega_{c})}][1-e^{-\mathcal C_q(\tilde \Omega, \tilde \omega_{c})}]}{e^{-\mathcal A_q(\tilde \Omega)}}\right)^{q} \nonumber \\
		& \times & \Bigg[ \frac{e^{-\mathcal A_q(\tilde \Omega)}}{[1-e^{-\mathcal B_q(\tilde \Omega, \tilde \omega_{c})}][1-e^{-\mathcal C_q(\tilde \Omega, \tilde \omega_{c})}]} \frac{e}{Mc\omega_c} \left(\frac{MV\omega_{c}}{2\pi\hbar}\right)\left(\frac{M}{2\pi\hbar^{2}\beta}\right)^{1/2}\Gamma\left(\frac{1}{q-1},\frac{1}{2}\right) \nonumber \\ & & + \left(\frac{MV\omega_{c}}{2\pi\hbar}\right)\left(\frac{M}{2\pi\hbar^{2}\beta}\right)^{1/2}\Gamma\left(\frac{1}{q-1},\frac{1}{2}\right) \times \frac{e^{-\mathcal A_q(\tilde \Omega) }}{[1-e^{-\mathcal B_q(\tilde \Omega, \tilde \omega_{c})}][1-e^{-\mathcal C_q(\tilde \Omega, \tilde \omega_{c})}]}\nonumber \\ & & \times  \left( -\frac{\partial \mathcal A_q(\tilde \Omega) }{\partial B} + \frac{e^{-\mathcal B_q(\tilde \Omega, \tilde \omega_{c})}}{1-e^{-\mathcal B_q(\tilde \Omega, \tilde \omega_{c})}}\frac{\partial \mathcal B_q(\tilde \Omega, \tilde \omega_{c})}{\partial B} + \frac{e^{-\mathcal C_q(\tilde \Omega, \tilde \omega_{c})}}{1-e^{-\mathcal C_q(\tilde \Omega, \tilde \omega_{c})}}\frac{\partial \mathcal C_q(\tilde \Omega, \tilde \omega_{c})}{\partial B} \right) \Bigg],
	\end{eqnarray}
}

where
{\small \begin{align}
		\mathcal A_q(\tilde \Omega) &= \left(\frac{3-q}{2}\right)\beta\hbar\tilde{\Omega}\left[\frac{1}{2}-\frac{e(E_{1}x_{0}+E_{2}y_{0})}{2\hbar\tilde{\Omega}}\right] \\ \mathcal B_q(\tilde \Omega, \tilde \omega_{c}) &= \left(\frac{3-q}{2}\right)\beta\hbar\tilde{\Omega}\left(1-\frac{\tilde{\omega_{c}}}{\tilde{\Omega}}\right), \quad \mathcal C_q(\tilde \Omega, \tilde \omega_{c}) = \left(\frac{3-q}{2}\right)\beta\hbar\tilde{\Omega}\left(1+\frac{\tilde{\omega_{c}}}{\tilde{\Omega}}\right),
	\end{align}
}	
with their corresponding derivatives supplied as below

{\small \begin{eqnarray}\label{qty000}
		\frac{\partial \mathcal A_q(\tilde \Omega)}{\partial B} & = \left(\frac{3-q}{2}\right)\beta\hbar \left[ -\frac{e\Omega\theta}{4c\sqrt{1-\frac{M\omega_{c}}{2}\theta + \left(\frac{M \Omega}{4}\theta\right)^{2}}} \left(\frac{1}{2}-\frac{e(E_{1}x_{0}+E_{2}y_{0})}{2\hbar\tilde{\Omega}}\right) \right. \cr & \quad \left. + \tilde{\Omega} \frac{e(E_{1}x_{0}+E_{2}y_{0})}{2\hbar\tilde{\Omega}^2} \left(-\frac{e\Omega\theta}{4c\sqrt{1-\frac{M\omega_{c}}{2}\theta + \left(\frac{M \Omega}{4}\theta\right)^{2}}}\right) \right],
	\end{eqnarray}
}
{\small \begin{eqnarray}\label{qty001}
		\frac{\partial \mathcal B_q(\tilde \Omega, \tilde \omega_{c})}{\partial B} & = \left(\frac{3-q}{2}\right)\beta\hbar \left[ -\frac{e\Omega\theta}{4c\sqrt{1-\frac{M\omega_{c}}{2}\theta + \left(\frac{M \Omega}{4}\theta\right)^{2}}} \left(1-\frac{\tilde{\omega}_{c}}{\tilde{\Omega}}\right) \right. \cr & \quad \left. - \tilde{\Omega} \frac{\tilde{\Omega}\frac{e}{Mc}\left(1 - \frac{2\omega^{2}_{0}}{\omega_{c}}M\theta\right) - \tilde{\omega}_c\left(-\frac{e\Omega\theta}{4c\sqrt{1-\frac{M\omega_{c}}{2}\theta + \left(\frac{M \Omega}{4}\theta\right)^{2}}}\right)}{\tilde{\Omega}^2} \right],
	\end{eqnarray}
}
{\small \begin{eqnarray}\label{qty003}
		\frac{\partial \mathcal C_q(\tilde \Omega, \tilde \omega_{c})}{\partial B} & = \left(\frac{3-q}{2}\right)\beta\hbar \left[ -\frac{e\Omega\theta}{4c\sqrt{1-\frac{M\omega_{c}}{2}\theta + \left(\frac{M \Omega}{4}\theta\right)^{2}}} \left(1+\frac{\tilde{\omega}_{c}}{\tilde{\Omega}}\right) \right. \cr & \quad \left. + \tilde{\Omega} \frac{\tilde{\Omega}\frac{e}{Mc}\left(1 - \frac{2\omega^{2}_{0}}{\omega_{c}}M\theta\right) - \tilde{\omega}_c\left(-\frac{e\Omega\theta}{4c\sqrt{1-\frac{M\omega_{c}}{2}\theta + \left(\frac{M \Omega}{4}\theta\right)^{2}}}\right)}{\tilde{\Omega}^2} \right].
	\end{eqnarray}
}

From the magnetization, we derive the magnetic susceptibility given by 
{\small \begin{eqnarray}
		\chi_{q}^{(\theta)} &=& -\frac{\partial M_{q}^{(\theta)}}{\partial B} = -\frac{\partial^2 F_{q}^{(\theta)}}{\partial B^2},
	\end{eqnarray}
}

and obtain from (\ref{qty000}) to (\ref{qty003}),

{\small \begin{eqnarray}\label{eq15}
		\chi_{q}^{(\theta)} &=& -\frac{\partial M_{q}^{(\theta)}}{\partial B} \cr &=& \frac{n(1-q)}{\beta} \left(\frac{MV\omega_{c}}{2\pi\hbar}\right)^{-q}\left(\frac{M}{2\pi\hbar^{2}\beta}\right)^{-q/2}\Gamma\left(\frac{1}{q-1},\frac{1}{2}\right)^{-q} \nonumber \\ &\times& \Bigg\{ -\frac{\partial}{\partial B}\left(\frac{[1-e^{-\mathcal B_q(\tilde \Omega, \tilde \omega_{c})}][1-e^{-\mathcal C_q(\tilde \Omega, \tilde \omega_{c})}]}{e^{-\mathcal A_q(\tilde \Omega)}}\right)^{q} \nonumber \\ &\times& \Bigg[ \frac{e}{Mc\omega_c} + \left( -\frac{\partial \mathcal A_q(\tilde \Omega)}{\partial B} + \frac{e^{-\mathcal B_q(\tilde \Omega, \tilde \omega_{c})}}{1-e^{-\mathcal B_q(\tilde \Omega, \tilde \omega_{c})}}\frac{\partial \mathcal B_q(\tilde \Omega, \tilde \omega_{c})}{\partial B} + \frac{e^{-\mathcal C_q(\tilde \Omega, \tilde \omega_{c})}}{1-e^{-\mathcal C_q(\tilde \Omega, \tilde \omega_{c})}}\frac{\partial \mathcal C_q(\tilde \Omega, \tilde \omega_{c})}{\partial B} \right) \Bigg] \nonumber \\ &+& \left(\frac{[1-e^{-\mathcal B_q(\tilde \Omega, \tilde \omega_{c})}][1-e^{-\mathcal C_q(\tilde \Omega, \tilde \omega_{c})}]}{e^{-\mathcal A_q(\tilde \Omega)}}\right)^{q} \nonumber \\ &\times& \frac{\partial}{\partial B} \Bigg[ \frac{e}{Mc\omega_c} + \left( -\frac{\partial \mathcal A_q(\tilde \Omega)}{\partial B} + \frac{e^{-\mathcal B_q(\tilde \Omega, \tilde \omega_{c})}}{1-e^{-\mathcal B_q(\tilde \Omega, \tilde \omega_{c})}}\frac{\partial \mathcal B_q(\tilde \Omega, \tilde \omega_{c})}{\partial B}\right.\cr 
		&&\left.+ \frac{e^{-\mathcal C_q(\tilde \Omega, \tilde \omega_{c})}}{1-e^{-\mathcal C_q(\tilde \Omega, \tilde \omega_{c})}}\frac{\partial \mathcal C_q(\tilde \Omega, \tilde \omega_{c})}{\partial B} \right) \Bigg] \Bigg\}.
	\end{eqnarray}
}

When $q\rightarrow 1$, the derived quantities in (\ref{eq13}), (\ref{eq14}), and (\ref{eq15}) are obtained as follows:
\begin{enumerate}
	\item
	{\small \begin{eqnarray}\label{frreeenrg000}
			F_{1}^{\left(\theta\right)} = \lim_{q \to 1} F_q^{(\theta)} &=& -\frac{n}{\beta} \ln\left(\frac{MV\omega_{c}}{2\pi\hbar}\sqrt{\frac{M}{2\pi\hbar^{2}\beta}}\right) + \frac{n\hbar\tilde{\Omega}}{2} - \frac{ne(E_{1}x_{0}+E_{2}y_{0})}{2} \cr && + \frac{n}{\beta} \ln\left[ \left(1-e^{-\beta\hbar(\tilde{\Omega}-\tilde{\omega_c})}\right)\left(1-e^{-\beta\hbar(\tilde{\Omega}+\tilde{\omega_c})}\right) \right];
		\end{eqnarray}
	}
	\item
	{\small \begin{eqnarray}
			M_{1}^{(\theta)} &=& \lim_{q \to 1} M_{q}^{(\theta)} = \frac{n}{\beta} \left(\frac{MV\omega_c}{2\pi\hbar}\right)^{-1}\left(\frac{M}{2\pi\hbar^2\beta}\right)^{-1/2}\times \frac{[1-e^{-\mathcal B_1(\tilde \Omega, \tilde \omega_{c})}][1-e^{-\mathcal C_1(\tilde \Omega, \tilde \omega_{c})}]}{e^{-\mathcal A_1(\tilde \Omega)}} \nonumber \\ &\times& \Bigg[ \frac{e}{Mc\omega_c} + \left( -\frac{\partial \mathcal A_1(\tilde \Omega)}{\partial B} + \frac{e^{-\mathcal B_1(\tilde \Omega, \tilde \omega_{c})}}{1-e^{-\mathcal B_1(\tilde \Omega, \tilde \omega_{c})}}\frac{\partial \mathcal B_1(\tilde \Omega, \tilde \omega_{c})}{\partial B}\right.\cr
			&&\left.+ \frac{e^{-\mathcal C_1(\tilde \Omega, \tilde \omega_{c})}}{1-e^{-\mathcal C_1(\tilde \Omega, \tilde \omega_{c})}}\frac{\partial \mathcal C_1(\tilde \Omega, \tilde \omega_{c})}{\partial B} \right) \Bigg]
		\end{eqnarray}
	}
	\item
	{\small \begin{eqnarray}
			\chi_{1}^{(\theta)} = \lim_{q \to 1} \chi_q^{(\theta)} &=& -\frac{\partial M_{1}^{(\theta)}}{\partial B}\nonumber\\
			&=& \frac{n}{\beta} \left(\frac{MV\omega_c}{2\pi\hbar}\right)^{-1}\left(\frac{M}{2\pi\hbar^2\beta}\right)^{-1/2} \nonumber \times \Bigg\{ -\frac{\partial}{\partial B}\left(\frac{[1-e^{-\mathcal B_1(\tilde \Omega, \tilde \omega_{c})}][1-e^{-\mathcal C_1(\tilde \Omega, \tilde \omega_{c})}]}{e^{-\mathcal A_1(\tilde \Omega)}}\right) \nonumber \\ &\times& \Bigg[ \frac{e}{Mc\omega_c} + \left( -\frac{\partial \mathcal A_1(\tilde \Omega)}{\partial B} + \frac{e^{-\mathcal B_1(\tilde \Omega, \tilde \omega_{c})}}{1-e^{-\mathcal B_1(\tilde \Omega, \tilde \omega_{c})}}\frac{\partial \mathcal B_1(\tilde \Omega, \tilde \omega_{c})}{\partial B}\nonumber \right.\\ 
			&&\left.+\frac{e^{-\mathcal C_1(\tilde \Omega, \tilde \omega_{c})}}{1-e^{-\mathcal C_1(\tilde \Omega, \tilde \omega_{c})}}\frac{\partial \mathcal C_1(\tilde \Omega, \tilde \omega_{c})}{\partial B} \right) \Bigg] \nonumber +\frac{[1-e^{-\mathcal B_1(\tilde \Omega, \tilde \omega_{c})}][1-e^{-\mathcal C_1(\tilde \Omega, \tilde \omega_{c})}]}{e^{-\mathcal A_1(\tilde \Omega)}} \nonumber \\ &\times& \frac{\partial}{\partial B} \Bigg[ \frac{e}{Mc\omega_c} + \left( -\frac{\partial \mathcal A_1(\tilde \Omega)}{\partial B} + \frac{e^{-\mathcal B_1(\tilde \Omega, \tilde \omega_{c})}}{1-e^{-\beta_1}}\frac{\partial \mathcal B_1(\tilde \Omega, \tilde \omega_{c})}{\partial B}\right.\nonumber\\
			&&\left. + \frac{e^{-\mathcal C_1(\tilde \Omega, \tilde \omega_{c})}}{1-e^{-\mathcal C_1(\tilde \Omega, \tilde \omega_{c})}}\frac{\partial \mathcal C_1(\tilde \Omega, \tilde \omega_{c})}{\partial B} \right) \Bigg] \Bigg\}.
		\end{eqnarray}
	}
	
	where
	{\small \begin{eqnarray}
			\mathcal A_1(\tilde \Omega) &=& \beta\hbar\tilde{\Omega}\left[\frac{1}{2}-\frac{e(E_1x_0+E_2y_0)}{2\hbar\tilde{\Omega}}\right]\cr \mathcal B_1(\tilde \Omega, \tilde \omega_{c}) &=& \beta\hbar\tilde{\Omega}\left(1-\frac{\tilde{\omega_c}}{\tilde{\Omega}}\right)\qquad \mathcal C_1(\tilde \Omega, \tilde \omega_{c}) = \beta\hbar\tilde{\Omega}\left(1+\frac{\tilde{\omega_c}}{\tilde{\Omega}}\right).
		\end{eqnarray} 
	}
\end{enumerate}

\subsection{Mathematical framework and validity analysis}

Statistical mechanics can be interpreted as a Hilhorst integral transformation of the corresponding standard quantities, based on the definition of the gamma function. To ensure the work's self-containment, this section establishes the rigorous mathematical foundation for our thermodynamic analysis within the framework of Tsallis nonextensive statistics, with a particular emphasis on the interplay between the nonextensivity parameter $q$ and the noncommutative parameter $\theta$.

\subsection{Validity domains and coupled conditions on $q$ and $\theta$}

The mathematical validity of our approach is governed by several fundamental requirements that couple the nonextensivity parameter $q$ with the noncommutative parameter $\theta$. Table~\ref{tab:validity} summarizes the domains of validity for the Hilhorst transformation applied to the noncommutative system.

\begin{table}[h]
	\centering
	\begin{tabular}{|l|c|c|}
		\hline
		\textbf{Domain} & \textbf{Parameter $q$} & \textbf{Conditions on $\theta$} \\
		\hline
		Main & $q \in ]1, +\infty[ \backslash \{3/2\}$ & $0 < M\omega_c\theta < 2$ \\
		\hline
		Restricted & $q \in (3/2, \infty)$ & $M\omega_c\theta < 1 + \frac{2}{q-1}$ \\
		\hline
		Critical & $q \in (0, 1)$ & $M\omega_c\theta < 1 - \frac{1}{2(1-q)}$ \\
		\hline
		\multicolumn{3}{|c|}{\textbf{Additional universal conditions}} \\
		\hline
		\multicolumn{3}{|l|}{$\beta \in (0, \infty)$, $T \in (0, \infty)$, $B > 0$, $E_1, E_2 \geq 0$} \\
		\hline
		\multicolumn{3}{|l|}{Positivity constraint: $\tilde{\Omega}^2 = \Omega^2\left[1-\frac{M\omega_c\theta}{2} + \left(\frac{M\Omega\theta}{4}\right)^2\right] > 0$} \\
		\hline
		\multicolumn{3}{|l|}{Convergence condition: $(q-1)\beta E_{\tilde{n}_+,\tilde{n}_-} < 1$ for all quantum states} \\
		\hline
	\end{tabular}
	\caption{Validity domains of the Hilhorst transformation for the noncommutative system with external electric fields. The coupling between $q$ and $\theta$ ensures the physical consistency of the nonextensive statistics in noncommutative space.}
	\label{tab:validity}
\end{table}

The transformation is valid for $q > 1$ (nonextensive regime), where the noncommutative Hamiltonian spectrum $H_{\theta}$ exhibits discrete energy levels bounded from below. The convergence of the integral in Eq.~(\ref{eq04}) is ensured for $\beta H_{\theta} > 0$, and the conditions on the gamma functions given in Eqs.~(\ref{eq03}-\ref{eq05}) must be satisfied simultaneously with the constraints on $\theta$.

\subsection{Physical domain of application}

The theoretical framework developed here applies within specific physical regimes where both noncommutativity and nonextensivity play significant roles:

\begin{itemize}
	\item \textbf{Finite temperatures}: where $\beta^{-1} = k_BT$ is well defined, with the additional constraint that thermal energy remains comparable to or larger than noncommutative corrections: $k_BT \gtrsim \hbar\omega_c M\omega_c\theta$;
	
	\item \textbf{Moderate magnetic fields}: where the cyclotron frequency $\omega_c$ satisfies $\omega_c < \frac{2}{M\theta}$ to ensure $\tilde{\Omega}^2 > 0$. This constraint becomes more restrictive for larger values of $q$ according to:
	\begin{eqnarray}
		B < B_{\text{max}}(q,\theta) = \frac{2Mc}{e\theta}\left(1 + \frac{\delta(q)}{(q-1)}\right)^{-1},
	\end{eqnarray}
	where $\delta(q)$ is a positive function of $q$ encoding the interplay between nonextensivity and noncommutativity;
	
	\item \textbf{Weak to moderate electric fields}: ensuring that the energy shift $E_0 = -\frac{e}{2\hbar\tilde{\Omega}}(E_1 x_0 + E_2 y_0)$ satisfies:
	\begin{eqnarray}
		|E_0| < \frac{1}{(q-1)\beta}\left(1 - M\omega_c\theta/2\right),
	\end{eqnarray}
	guaranteeing that the nonextensive corrections remain physically meaningful and mathematically tractable;
	
	\item \textbf{Coupled noncommutative-nonextensive regime}: the parameter space must satisfy the joint constraint:
	\begin{eqnarray}\label{eq_coupled_constraint}
		(q-1)\beta\hbar\tilde{\Omega} < 1 \quad \text{and} \quad M\omega_c\theta < 2\left(1 - \frac{1}{(q-1)\beta\hbar\Omega}\right).
	\end{eqnarray}
	
	This ensures that both the $q$-exponential deformation and the noncommutative geometry corrections are well-controlled and mutually consistent.
\end{itemize}

These conditions ensure that the nonextensive corrections remain physically meaningful 

and mathematically tractable throughout the parameter space of interest.

\subsection{Constraints on the nonextensivity parameter $q$ and noncommutative parameter $\theta$}

The convergence of the gamma function $\Gamma\left(\frac{1}{q-1}\right)$, as required by Eqs.~(\ref{eq03}--\ref{eq05}), imposes the fundamental constraint:
\begin{eqnarray}\label{eq_constraint_q}
	\alpha = \frac{1}{q-1} > 0 \quad \Rightarrow \quad q > 1.
\end{eqnarray}

However, this condition alone is insufficient when considering the noncommutative structure. The effective frequencies $\tilde{\Omega}$ and $\tilde{\omega}_c$ must remain real and positive, imposing additional constraints on the allowed region in the $(q,\theta)$ parameter space.

\paragraph{Constraint from positivity of $\tilde{\Omega}$:}
The effective frequency $\tilde{\Omega}$ must satisfy $\tilde{\Omega}^2 > 0$, which requires:
\begin{eqnarray}\label{eq_theta_bound1}
	\frac{M\omega_c\theta}{2} - \left(\frac{M\Omega\theta}{4}\right)^2 < 1.
\end{eqnarray}

For small $\theta$ (perturbative regime), this simplifies to $M\omega_c\theta < 2$.

\paragraph{Constraint from thermodynamic stability:}
The partition function $Z_q^{(\theta)}$ must converge, requiring that the argument of the $q$-exponential remains finite. This imposes:
\begin{eqnarray}\label{eq_coupled_qtheta}
	(q-1)\beta\hbar\tilde{\Omega}\left[1 + \frac{\tilde{\omega}_c}{\tilde{\Omega}}\right] < 2.
\end{eqnarray}

Expanding to first order in $\theta$, (\ref{eq_coupled_qtheta}) yields a coupled constraint:
\begin{eqnarray}\label{eq_explicit_coupled}
	q < 1 + \frac{2}{\beta\hbar\Omega\left[2 - M\omega_c\theta + \mathcal{O}(\theta^2)\right]}.
\end{eqnarray}

Equation (\ref{eq_explicit_coupled}) demonstrates that (larger noncommutativity (larger $\theta$) restricts the allowed range of nonextensivity (smaller maximum $q$)), revealing a fundamental trade-off between these two types of generalized statistics.

\paragraph{Physical interpretation:}
The coupled constraint in Eq.~(\ref{eq_explicit_coupled}) has deep physical meaning:
\begin{itemize}
	\item in the commutative limit ($\theta \to 0$): the constraint reduces to the standard Tsallis condition $q < 1 + \frac{1}{\beta\hbar\Omega}$;
	
	\item for finite $\theta$: noncommutative geometry corrections effectively reduces the accessible $q$-space, suggesting that spatial noncommutativity and statistical nonextensivity are complementary rather than additive modifications;
	
	\item at strong noncommutativity ($M\omega_c\theta \to 2$): the system approaches $q \to 1$ (extensive limit), indicating that strong noncommutative effects suppress nonextensive behavior.
\end{itemize}

For $q < 1$, an alternative regularization procedure must be employed, with modified 

constraints:
{\small \begin{eqnarray}\label{eq_q_less_1}
		M\omega_c\theta < 1 - \frac{1}{2(1-q)\beta\hbar\Omega},
	\end{eqnarray}
}

leading to the modified expression:
{\small \begin{eqnarray}
		\frac{\Gamma\left(\frac{1}{1-q}+1\right)}{(1-q)^{\frac{1}{2}}\Gamma\left(\frac{1}{1-q}+\frac{3}{2}\right)}, \quad \text{for } q < 1,
\end{eqnarray}}

as shown in Eq.~(\ref{eq07}).
\subsection{Thermodynamic consistency conditions}

The mathematical transformation requires that the change of variable in Eq.~(\ref{eq05}) satisfies:
{\small \begin{eqnarray}\label{eq_consistency}
		t = \nu\left[1-\left(1-q\right)\beta H_{\theta}\right] > 0,
	\end{eqnarray}
}

for each quantum state characterized by the energy eigenvalues $E_{\tilde{n}_+, \tilde{n}_-}$ in (\ref{specteig000}). The 

thermodynamic consistency demands:
{\small \begin{eqnarray}\label{eq_consistency2}
		\left[1 - (1-q)\beta E_{\tilde{n}_+, \tilde{n}_-}\right] > 0 \quad \text{for all } (\tilde{n}_+, \tilde{n}_-),
	\end{eqnarray}
}

where the energy spectrum (\ref{specteig000}) can be recast as follows:
{\small \begin{eqnarray}\label{eq_spectrum}
		E_{\tilde{n}_+, \tilde{n}_-} = \hbar\tilde{\Omega}_+\tilde{n}_+ + \hbar\tilde{\Omega}_-\tilde{n}_- + E_0,
	\end{eqnarray}
}

with $\tilde{n}_{\pm} \in \mathbb{N}$, and
{\small \begin{eqnarray}
		\tilde{\Omega}_{\pm} = \frac{1}{2\tilde{\Omega}}\left(\tilde{\Omega} \pm \tilde{\omega}_c\right), \qquad E_0 = \frac{\hbar\tilde{\Omega}}{2} - \frac{e}{2\hbar\tilde{\Omega}}(E_1 x_0 + E_2 y_0).
	\end{eqnarray}
}

Since the spectrum is unbounded from above (as $\tilde{n}_{\pm} \to \infty$), the condition (\ref{eq_consistency2}) can only 

be satisfied 

if:
{\small \begin{eqnarray}\label{eq_ground_state_condition}
		(q-1)\beta E_{\text{ground}} < 1, \quad \text{where } E_{\text{ground}} = E_{0,0} = E_0,
	\end{eqnarray}
}

and the series in Eq.~(\ref{equaint007}) converges. This imposes the fundamental constraint linking $q$,

 $\theta$, and the electric field parameters:
{\small \begin{eqnarray}\label{eq_full_constraint}
		(q-1)\beta\hbar\tilde{\Omega}\left[\frac{1}{2} - \frac{e}{2\hbar^2\tilde{\Omega}^2}(E_1 x_0 + E_2 y_0)\right] < 1.
	\end{eqnarray}
}

These constraints ensure that the $q$-exponential functions remain well-defined and physi-

cally meaningful throughout the entire spectrum of the noncommutative Hamiltonian

 $H_{\theta}$. From Eq.~(\ref{eq04}), the partition function $Z_{q}^{(\theta)}$ expresses, by setting $x=\beta\hbar\tilde{\Omega}$ and 
 
 considering the cylinder of height $L_{z}$ and radius $R$, as (\ref{eq09})

\subsection{Convergence analysis and series summation}

The mathematical rigor of our approach is further validated by examining the convergence properties of the transformed partition function. From the following identity \cite{bienvenu and aremua}
{\small \begin{eqnarray}\label{eq_gamma_identity}
		\Gamma\left(a,b\right)=\frac{\Gamma\left(a-b\right)}{\Gamma(a)}e^{b\log(a)}, \quad a = \frac{1}{q-1}, \, b = \frac{1}{2},
	\end{eqnarray}
}

we get
{\small\begin{eqnarray}\label{eq11}
		\Gamma\left(\frac{1}{q-1}-\frac{1}{2}\right)&=&\Gamma\left(\frac{1}{q-1},\frac{1}{2}\right)\Gamma\left(\frac{1}{q-1}\right)\left(q-1\right)^{\frac{1}{2}}.
	\end{eqnarray}
}

The Hilhorst method, when applied to the Tsallis framework with noncommutative corrections, maintains the essential thermodynamic properties while extending the validity domain beyond the classical Boltzmann-Gibbs regime. The convergence of the double sum in Eq.~(\ref{equaint007}) is guaranteed by the coupled constraints on $q$ and $\theta$ established in Eqs.~(\ref{eq_coupled_qtheta})-(\ref{eq_full_constraint}). This extension is particularly crucial for understanding systems where standard extensive thermodynamics fails to capture the underlying physics, such as systems with noncommutative spatial coordinates or those exhibiting long-range correlations.

\subsection{Recovery of standard statistical mechanics}

When $q\rightarrow 1$, we have $a\rightarrow\infty$, and $\displaystyle{\lim_{a\rightarrow+\infty}\Gamma\left(a,b\right)=1}$. Thereby, Eq.~(\ref{eq06}) recovers the standard partition function within the Boltzmann-Gibbs framework, that is (\ref{partidef000}), this limiting behavior confirms the consistency of our generalized approach and demonstrates that the nonextensive framework smoothly reduces to the standard statistical mechanics in the extensive limit $q \to 1$. Crucially, the presence of noncommutative parameters $\theta$, $x_0$, and $y_0$ persists even in this limit, reflecting the fundamental modification of the underlying phase space geometry induced by spatial noncommutativity. This demonstrates that noncommutativity and nonextensivity are independent generalizations that can be studied separately or in combination.

\subsection{Physical interpretation: the $(q,\theta)$-parameter space}

The coupled constraints on the nonextensivity parameter $q$ and the noncommutative parameter $\theta$ define a physically accessible region in the $(q,\theta)$-plane, which encodes several physically significant features:

\paragraph{Classification of physical regimes:}
\begin{itemize}
	\item \textbf{standard regime} ($q = 1$, $\theta = 0$): standard Boltzmann-Gibbs statistics in commutative space conventional quantum mechanics \cite{Landau1980StatisticalPhysics};
	
	\item \textbf{pure nonextensive regime} ($q > 1$, $\theta \approx 0$): systems with sub-additive entropy, characteristic of long-range interactions, without significant spatial noncommutativity. Relevant for gravitational systems, plasma physics \cite{Tsallis2009IntroductionNonextensive};
	
	\item \textbf{pure noncommutative regime} ($q \approx 1$, $\theta > 0$): systems in noncommutative space described by extensive statistics. Relevant for quantum Hall systems, strong magnetic field limits \cite{Douglas2001NoncommutativeField};
	
	\item \textbf{coupled regime} ($q > 1$, $\theta > 0$): combined effects of statistical nonextensivity and spatial noncommutativity. The constraint Eq.~(\ref{eq_explicit_coupled}) reveals an {\it{inverse correlation}}: stronger\\
	noncommutativity which restricts the degree of nonextensivity, suggesting a complementarity principle between these two generalizations \cite{Douglas2001NoncommutativeField};
	
	\item \textbf{critical boundary} ($q = 3/2$, arbitrary $\theta$): marks a transition in the convergence properties of the gamma function representation, where special care must be taken in the mathematical treatment \cite{Tsallis2009IntroductionNonextensive};
	
	\item \textbf{super-extensive regime} ($q < 1$, constrained $\theta$): systems with super-additive entropy, relevant for systems with bounded phase space or strong constraints, with modified noncommutative bounds given by Eq.~(\ref{eq_q_less_1}).
\end{itemize}

\paragraph{Physical observables depending on $(q,\theta)$:}

The interplay between $q$ and $\theta$ manifests in observable quantities:
\begin{itemize}
	\item \textbf{effective temperature}: $T_{\text{eff}}(q,\theta) = T\left[1 + (q-1)\langle E_{\theta}\rangle/k_BT\right]^{-1}$ where $\langle E_{\theta}\rangle$ depends on $\theta$ through $\tilde{\Omega}$ and $\tilde{\omega}_c$ \cite{Tsallis2009IntroductionNonextensive, Abe2000GeneralizedMolecular};
	
	\item \textbf{magnetic susceptibility}: exhibits enhanced diamagnetism with increasing $q$ and modified oscillations (noncommutative corrections) with increasing $\theta$;
	
	\item \textbf{heat capacity}: shows non-monotonic behavior in the $(q,\theta)$-plane, with possible peaks near the constraint boundary.
\end{itemize}

\paragraph{Complementarity principle:}
constraint Eq.~(\ref{eq_explicit_coupled}) suggests a {\it complementarity between nonextensivity and noncommutativity}: systems cannot simultaneously exhibit strong nonextensive behavior (large $q$) and strong spatial noncommutativity (large $\theta$). This has profound implications:
\begin{itemize}
	\item at the Planck scale, where $\theta$ may be maximal, statistical distributions may be forced toward extensivity ($q \to 1$);
	
	\item conversely, in astrophysical systems with strong nonextensive effects ($q \gg 1$),
	
	spatial noncommutativity must be minimal ($\theta \to 0$);
	
	\item intermediate mesoscopic systems can explore moderate values of both parameters.
\end{itemize}

The interplay between the nonextensivity parameter $q$ and the noncommutative parameter $\theta$ provides a rich framework for exploring quantum systems in modified phase space geometries, with potential applications ranging from quantum Hall systems to mesoscopic physics, quantum information theory, and potentially quantum gravity phenomenology \cite{Tsallis2009IntroductionNonextensive, Douglas2001NoncommutativeField}.

\section{Analysis and discussions on the derived thermodynamical quantities}

\begin{figure}[H]
	\centering
	\includegraphics[width=0.5\linewidth]{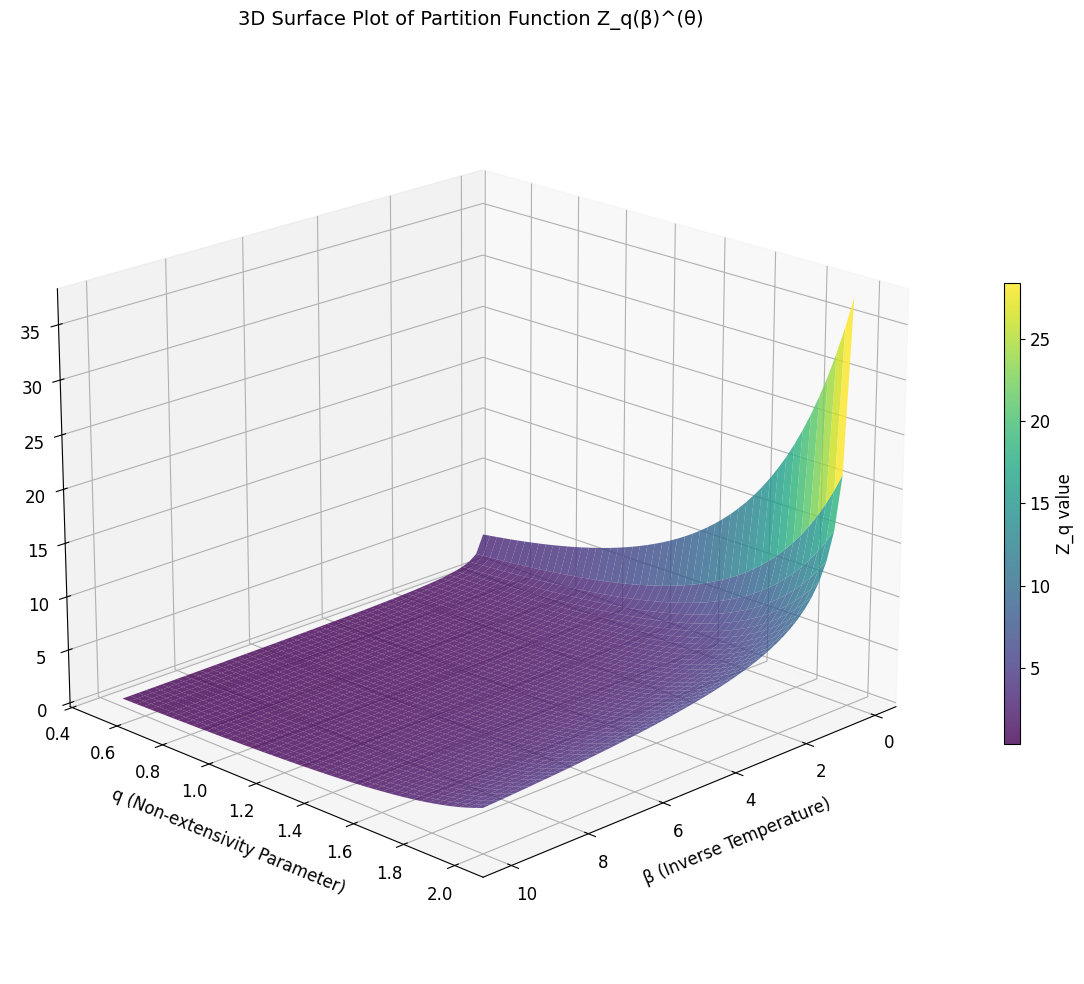}
	\caption{3D surface plot of the partition function \( Z_q(\beta) \) for an electron in a noncommutative plane. The surface shows the variation of \( Z_q \) as a function of the non-extensivity parameter \( q \) and the inverse temperature \( \beta \).}
	\label{fig:partition}
\end{figure}

\begin{figure}[H]
	\centering
	\includegraphics[width=0.5\textwidth]{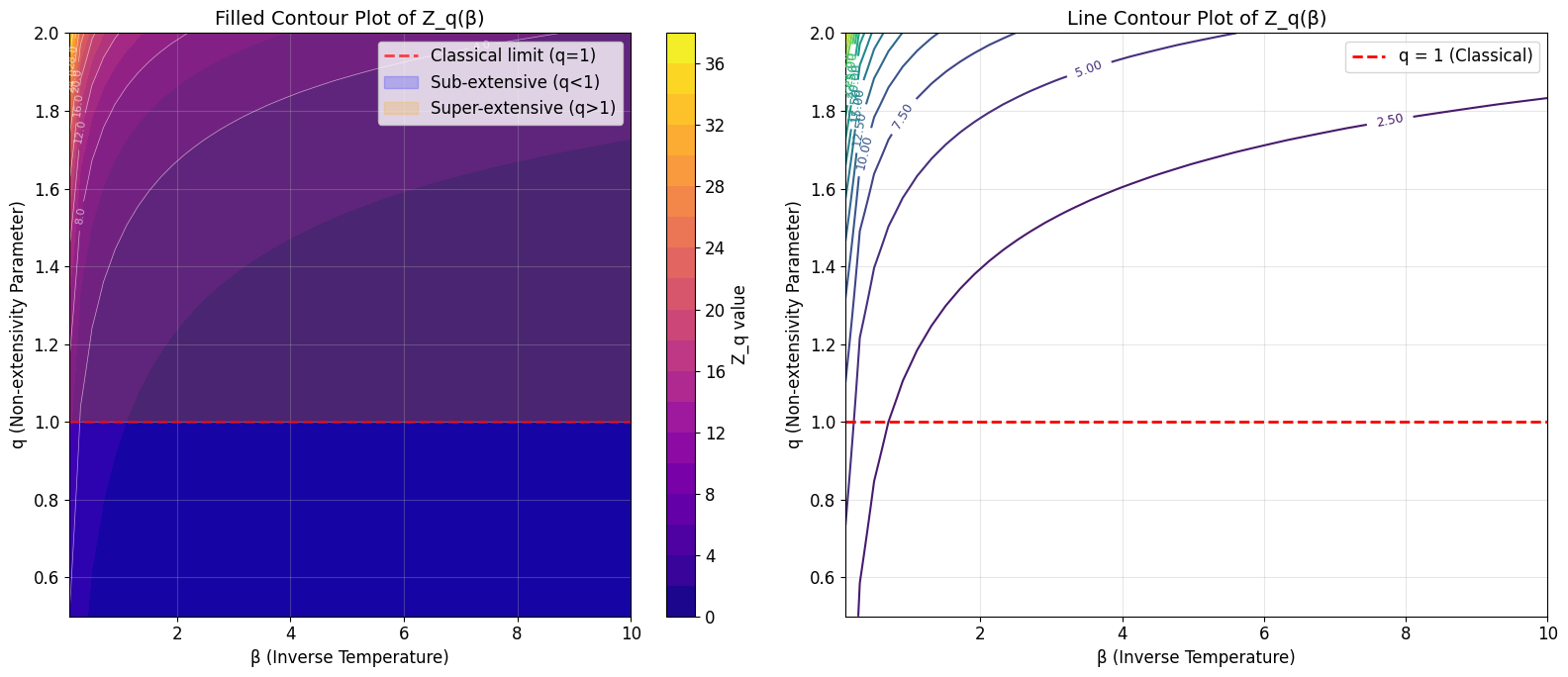}
	\caption{Contour plots of $Z_q(\beta)$: filled contours (left) and line contours (right). The dashed red line indicates the Boltzmann-Gibbs limit ($q=1$).}
	\label{fig:Zq}
\end{figure}

\begin{figure}[H]
	\centering
	\includegraphics[width=0.5\textwidth]{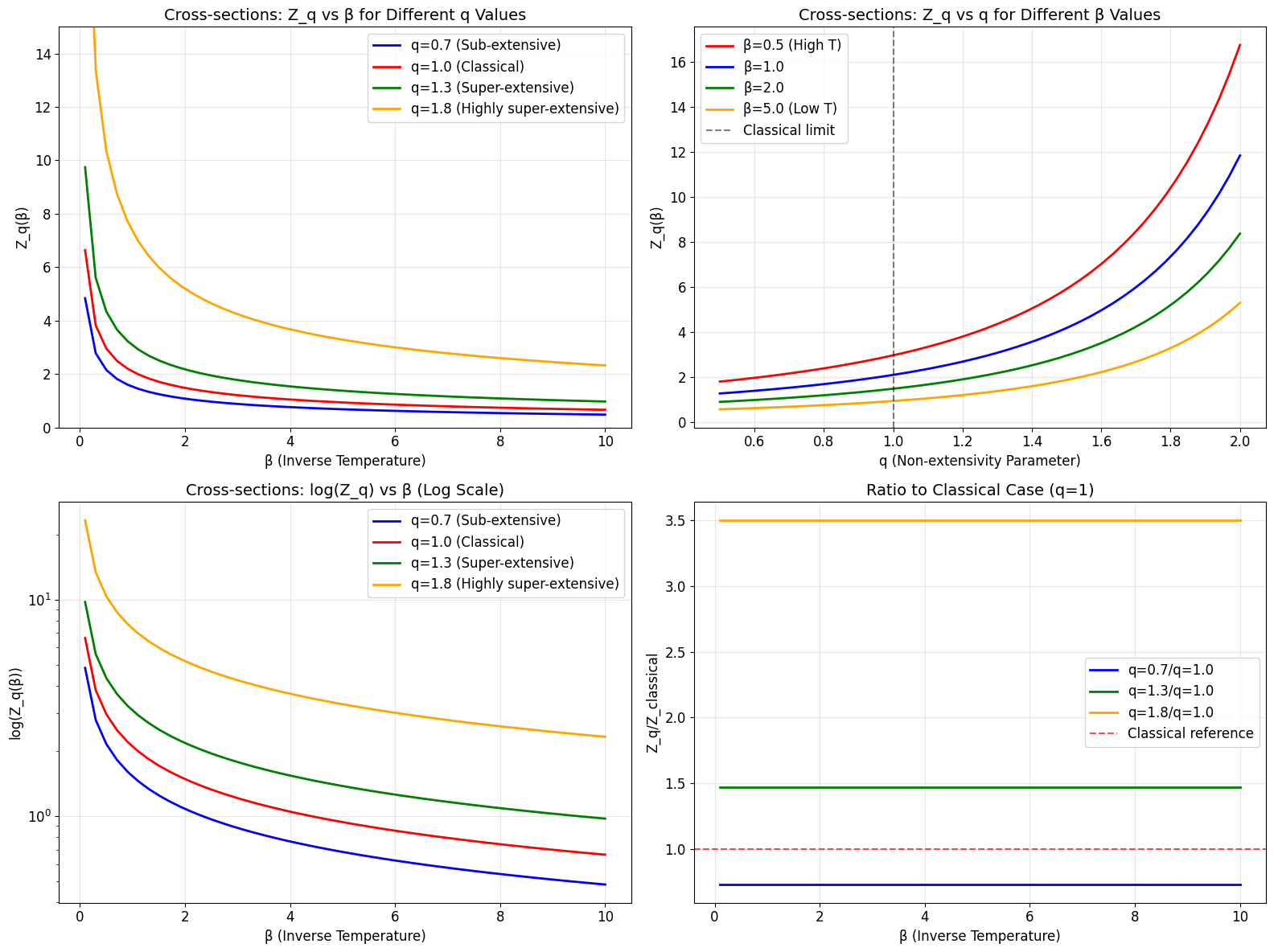}
	\caption{Cross-sectional analysis of $Z_q(\beta)$: 
		(top left) $Z_q$ vs.\ $\beta$ for different $q$,
		(top right) $Z_q$ vs.\ $q$ for different $\beta$,
		(bottom left) $\log(Z_q)$ vs.\ $\beta$,
		(bottom right) $Z_q/Z_{q=1}$ ratio vs.\ $\beta$.}
	\label{fig:Pq_section}
\end{figure}

\begin{figure}[H]
	\centering
	\includegraphics[width=0.5\textwidth]{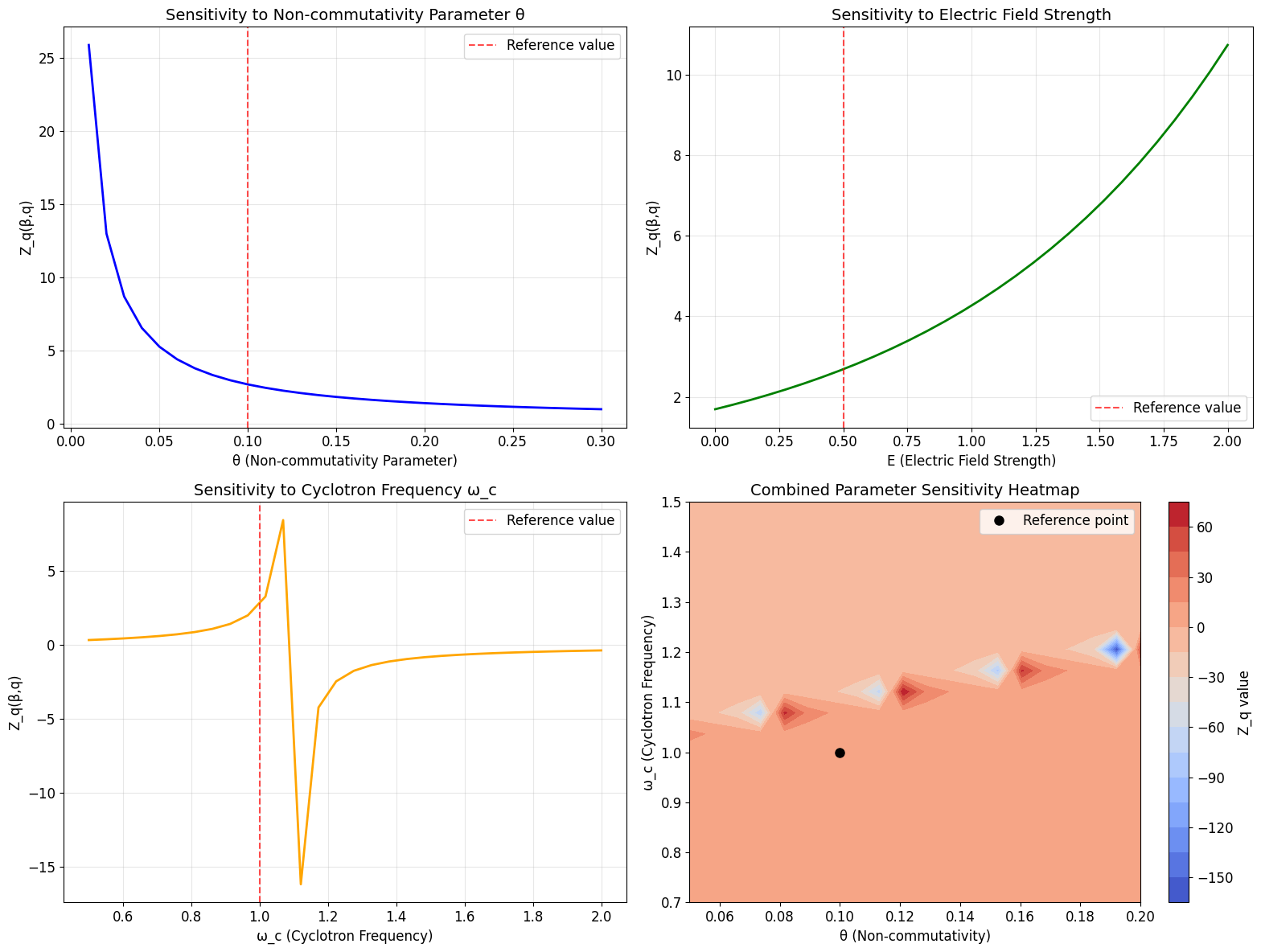}
	\caption{Sensitivity of $Z_q(\beta)$ to external parameters: 
		(top left) $\theta$ dependence,
		(top right) $E$ dependence,  
		(bottom left) $\omega_c$ dependence,
		(bottom right) $\theta$-$\omega_c$ joint sensitivity.
		Reference values: red dashed lines/black markers.}
	\label{fig:Zq_sensitivity}
\end{figure}

\begin{figure}[H]
	\centering
	\includegraphics[width=0.5\textwidth]{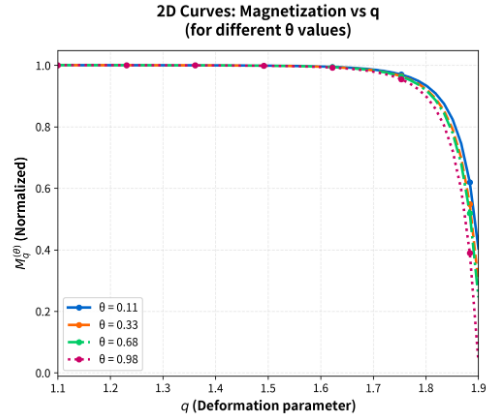}
	\caption{Dependence of normalized magnetization $M_{q}^{(\theta)}$ on the deformation parameter $q$ for different orientation angles $\theta$ in a two-dimensional $q$-deformed spin system.}
	\label{fig:Mq_theta_dependence}
\end{figure}

\begin{figure}[H]
	\centering
	\includegraphics[width=0.8\textwidth]{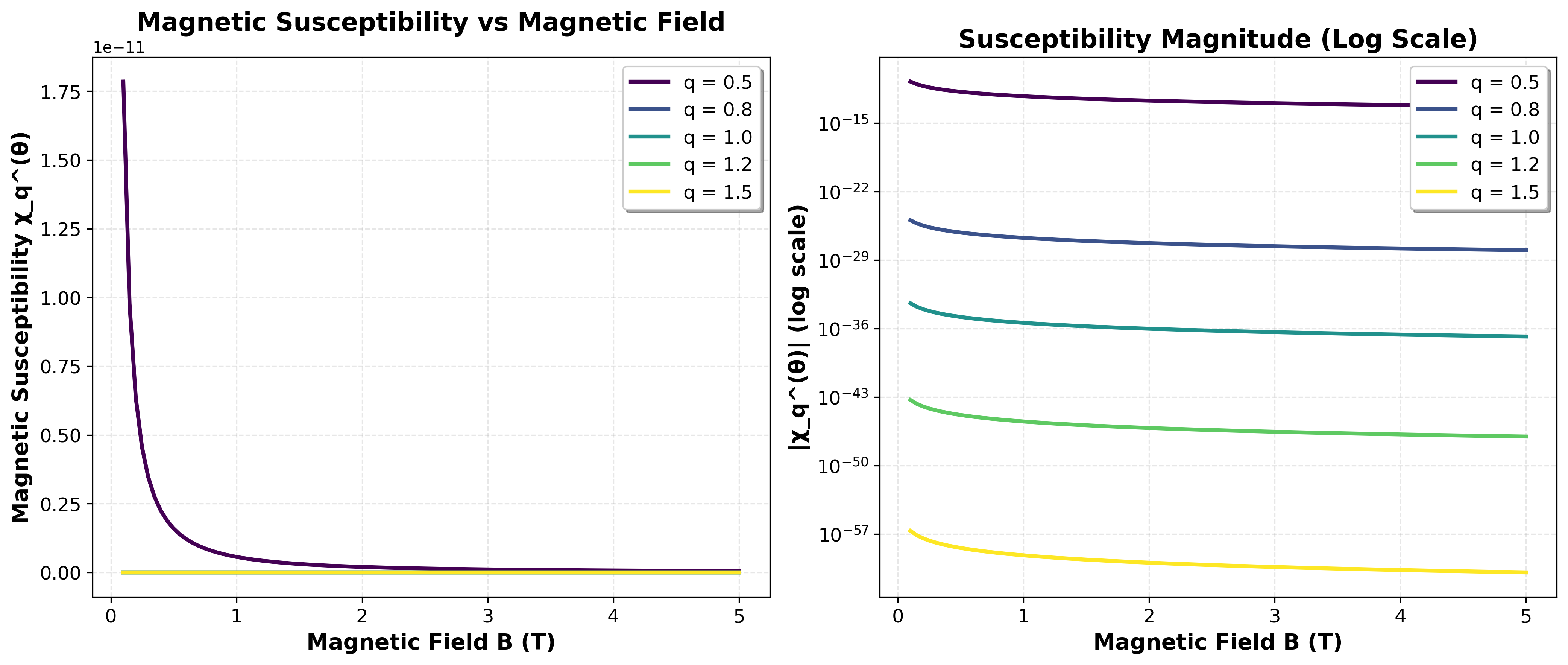}
	\caption{Magnetic susceptibility $\chi_{q}^{(\theta)}$ vs.\ $B$ for various $q$ and $\theta$. 
		Log scale shows field-induced suppression (magnetic saturation). 
		$\theta$-dependence reflects anisotropy; $q$-dependence indicates quantum deformation effects.}
	\label{fig:susceptibility_field}
\end{figure}

\subsection{Discussions and interpretations}

\begin{itemize}
	\item Figure~\ref{fig:partition} presents a three-dimensional visualisation of the quantum partition function \( Z_q(\beta) \), derived from equation (\ref{eq07}) of the source, within the framework of studying the thermodynamics of an electron in a noncommutative plane. This analysis of the parameter space \((q, \beta, Z_q(\beta))\) reveals the fundamental thermodynamic character of an electron in a noncommutative plane. The nonextensivity parameter \(q\) (X-axis) quantifies the deviation from standard Boltzmann-Gibbs statistics, where \(q = 1\) represents the commutative regime. A deviation from unity, induced by the spatial noncommutation relation \( [\hat{x}, \hat{y}] = i\theta \), signals the emergence of nonextensivity and long-range correlations \cite{Tsallis1988}. The inverse temperature \(\beta\) (Y-axis) is a standard thermodynamic variable, while the partition function \(Z_q(\beta)\) (Z-axis) encapsulates the system's full thermodynamic information. The most striking feature of the resulting surface is a sharp "cliff" near \(q = 1\), which visually represents a phase transition or crossover between the extensive (\(q = 1\)) and nonextensive (\(q \neq 1\)) regimes. This abrupt change highlights the non-perturbative nature of noncommutativity; an infinitesimal deviation from commutativity leads to a drastic restructuring of the thermodynamic landscape, underscoring that it is a fundamental modification of spacetime. Furthermore, the distinct behaviors for \(q < 1\) and \(q > 1\) suggest different physical phases, and the smooth variation with \(\beta\) confirms the expected temperature dependence. In conclusion, thermodynamics is inherently nonextensive, with the parameter \(\theta\) acting as the source for \(q\). The singular commutative limit (\(\theta \rightarrow 0\)) proves that noncommutative thermodynamics cannot be obtained as a continuous deformation of the standard case, necessitating a description within the framework of Tsallis statistics \cite{TSALLIS200947}.
	
	\item Figure~\ref{fig:Zq} presents two types of contour plots (filled and line) of the generalised partition function $Z_{q}(\beta)$ as a function of the inverse temperature $\beta$ (horizontal axis) and the non-extensivity parameter $q$ (vertical axis). This analysis is conducted in the framework of Tsallis statistics, which generalises the Boltzmann-Gibbs (BG) case \cite{Tsallis1988}. The contour plots of the partition function \(Z_{q}(\beta)\) delineate the thermodynamic landscape of the noncommutative electron. The filled contour (left) displays the \(Z_q\) magnitude using a colour gradient, while the line contours (right) indicate its levels. The classical Boltzmann-Gibbs limit at \(q=1\) (red dashed line) separates two distinct regimes: a \textbf{sub-extensive} phase (\(q<1\)) where correlations suppress \(Z_q\), and a \textbf{super-extensive} phase (\(q>1\)) where long-range interactions enhance it . The thermal dependence along \(\beta\) is standard, with \(Z_q\) decreasing at low temperatures. The stark contrast across \(q=1\) demonstrates the singular nature of the commutative limit and establishes Tsallis statistics as essential for describing the system's inherent non-extensivity \cite{Tsallis1988}.
	
	\item Figure~\ref{fig:Pq_section} illustrates different cross-sections of the generalised partition function $Z_q(\beta)$ in the framework of Tsallis statistics \cite{Tsallis1988}. The panels separately explore the dependence of $Z_q$ on the inverse temperature $\beta$ and on the non-extensivity parameter $q$, highlighting deviations from the Boltzmann-Gibbs limit ($q=1$). The dependence of the partition function \(Z_q\) on inverse temperature \(\beta\) for different \(q\) values demonstrates the core impact of non-extensivity. While \(Z_q\) decreases monotonically with \(\beta\) for all cases, its magnitude is strongly ordered by \(q\), with super-extensive values (\(q>1\)) enhancing the statistical weight of states and sub-extensive values (\(q<1\)) suppressing it \cite{Tsallis1988}. This \(q\)-dependence is most pronounced at high temperature (\(\beta \to 0\)) and diminishes at low temperature as the ground state dominates. The ratio \(Z_q/Z_{q=1}\) (bottom right), which is independent of \(\beta\), quantifies this deviation: it is below unity for \(q<1\) and above unity for \(q>1\). These results confirm \(q\) as a direct measure of deviation from Boltzmann-Gibbs statistics, with sub-extensivity (\(q<1\)) characterising constrained systems and super-extensivity (\(q>1\)) describing systems with long-range interactions or anomalous fluctuations.
	
	\item Figure~\ref{fig:Zq_sensitivity} characterises the parameter sensitivity of the generalised partition function \(Z_q(\beta, q)\) in a nonextensive, noncommutative framework. The analysis reveals distinct responses: \(Z_q\) decreases with the noncommutativity parameter \(\theta\), indicating a reduction in effective phase-space volume. Conversely, it increases monotonically with the electric field strength \(E\), which enlarges the energy spectrum and accessible states \cite{Tsallis1988}. The cyclotron frequency \(\omega_c\) induces non-linear, resonant-like variations due to Landau-level quantisation \cite{Curado1991}. The combined sensitivity map (bottom right) synthesises these effects, showing regions of enhanced (red) and suppressed (blue) \(Z_q\) under simultaneous variation of $\theta$ and $\omega_c$, highlighting critical interplay in noncommutative quantum statistical mechanics.
	
	\item Figure~\ref{fig:Mq_theta_dependence} presents the dependence of normalized magnetization $M_{q}^{(\theta)}$ on the deformation parameter $q$ reveals fundamental aspects of magnetic ordering in two-dimensional $q$-deformed spin systems. As shown in Fig.~1, the magnetization exhibits systematic suppression with increasing $q$ across all orientation angles $\theta$, indicating that algebraic deformation enhances quantum fluctuations that progressively disrupt magnetic order. The distinct $\theta$-dependence further demonstrates significant magnetic anisotropy, with the angular variation reflecting orientation-dependent energy landscapes. The monotonic decay suggests the approach toward a potential quantum critical point at $q_c > 1.9$, where long-range order may be fully suppressed. These findings highlight the intricate relationship between non-commutative algebraic structures and emergent magnetic phenomena in low-dimensional quantum materials.
	
	\item The magnetic susceptibility $\chi_{q}^{(\theta)}$ as a function of the applied magnetic field $B$ reveals key features of the magnetic response in the $q$-deformed system. As shown in Fig.~\ref{fig:susceptibility_field}, the susceptibility decreases with increasing magnetic field, following a characteristic trend toward magnetic saturation. This behavior is consistent with the suppression of spin fluctuations under strong external fields. The variation with angle $\theta$ underscores the anisotropic nature of the system, where different orientations exhibit distinct susceptibility profiles due to directional dependencies in the spin interactions. Furthermore, the influence of the deformation parameter $q$ highlights how quantum group deformations alter the magnetic properties, potentially enhancing or suppressing susceptibility depending on the specific $q$-value. These results emphasize the interplay between external magnetic fields, structural anisotropy, and algebraic deformation in determining the overall magnetic behavior of low-dimensional quantum systems \cite{Tsallis1988}.
\end{itemize}

\subsection{Discussions at high and low temperatures}

This paragraph is devoted to the analysis of the calculated thermal quantities' behavior. We first consider the asymptotic expressions at low temperatures, and then at high temperatures. In the sequel, we use the partition function expression supplied by (\ref{partidef000}), with $\Omega=\sqrt{\omega_{c}^{2}+4\omega_{0}^{2}}$ and $x = \hbar\beta\tilde{\Omega}$ the normalized inverse temperature parameter, and also the free energy given by (\ref{frreeenrg000}). We analyze these quantity expressions at high and low temperature limits, respectively.

\begin{itemize}
	\item {\bf Case of high temperature limit ($T \to \infty$, $x \to 0$).}
	
	In this context, since $x \ll 1$, we perform Taylor expansions of all exponential terms in (\ref{partidef000}). Therefore, we obtain the following approximations:
	\begin{enumerate}
		\item for the numerator:
		\begin{equation}
			e^{-x\left[\frac{1}{2}-\frac{e}{2\hbar\tilde{\Omega}}\left(E_{1}x_{0}+E_{2}y_{0}\right)\right]} \approx 1 - x\left[\frac{1}{2}-\frac{e}{2\hbar\tilde{\Omega}}\left(E_{1}x_{0}+E_{2}y_{0}\right)\right];
		\end{equation}
		\item for the denominator, using the result $1-e^{-y} \approx y$ when $y \ll 1$, we get
		\begin{equation}
			\left[1-e^{-\frac{x}{2}\left(1-\frac{\tilde{\omega}_{c}}{\tilde{\Omega}}\right)}\right] \approx \frac{x}{2}\left(1-\frac{\tilde{\omega}_{c}}{\tilde{\Omega}}\right), \quad \left[1-e^{-\frac{x}{2}\left(1+\frac{\tilde{\omega}_{c}}{\tilde{\Omega}}\right)}\right] \approx \frac{x}{2}\left(1+\frac{\tilde{\omega}_{c}}{\tilde{\Omega}}\right),
		\end{equation}
		such that, their product provides
		\begin{equation}
			\frac{x^2}{4}\left(1-\frac{\tilde{\omega}_{c}^2}{\tilde{\Omega}^2}\right) = \frac{x^2}{4}\frac{\tilde{\Omega}^2-\tilde{\omega}_{c}^2}{\tilde{\Omega}^2}.
		\end{equation}
	\end{enumerate}
	
	Substituting $x = \hbar\beta\tilde{\Omega}$ and simplifying, the partition function now reads:
	{\small \begin{equation}\label{approx000}
			Z_{1}^{(\theta)} \approx \frac{4VM\omega_{c}}{2\pi\hbar^4\beta^{5/2}(\tilde{\Omega}^2-\tilde{\omega}_{c}^2)}\left(\frac{M}{2\pi}\right)^{\frac{1}{2}}.
		\end{equation}
	}
	
	The approximation (\ref{approx000}) exhibits the characteristic classical behavior $Z \propto T^{5/2}$ as deduced above, where the effective frequency determines the density of states. Simplifying the logarithmic terms, we finally obtain for the free energy (\ref{frreeenrg000}):
	\begin{align}
		F_{1}^{(\theta)} &\approx -\frac{n}{\beta} \ln\left(\frac{MV\omega_{c}}{2\pi\hbar}\right) - \frac{n}{2\beta} \ln\left(\frac{M}{2\pi\hbar^{2}\beta}\right) \nonumber \\ &\quad + \frac{n\hbar\tilde{\Omega}}{2} - \frac{ne(E_{1}x_{0}+E_{2}y_{0})}{2} + \frac{2n}{\beta} \ln(\beta) + \frac{n}{\beta} \ln[\hbar^2(\tilde{\Omega}^2-\tilde{\omega}_c^2)].
	\end{align}
	
	In the high-temperature regime where \( \beta\hbar\tilde{\Omega} \ll 1 \), we employ the following Taylor expansions:
	
	{\small \begin{eqnarray}
			e^{-\mathcal B_1(\tilde \Omega, \tilde \omega_{c})}&\approx& 1 - \mathcal B_1(\tilde \Omega, \tilde \omega_{c}), \quad e^{-\mathcal C_1(\tilde \Omega, \tilde \omega_{c})} \approx 1 - \mathcal C_1(\tilde \Omega, \tilde \omega_{c}),\cr
			[1-e^{-\mathcal B_1(\tilde \Omega, \tilde \omega_{c})}][1-e^{-\mathcal C_1(\tilde \Omega, \tilde \omega_{c})}]&\approx& \mathcal B_1(\tilde \Omega, \tilde \omega_{c}) \mathcal C_1(\tilde \Omega, \tilde \omega_{c}) = (\beta\hbar\tilde{\Omega})^2 \left(1 - \frac{\tilde{\omega_c}^2}{\tilde{\Omega}^2}\right),\cr 
			\frac{e^{-\mathcal B_1(\tilde \Omega, \tilde \omega_{c})}}{1-e^{-\mathcal B_1(\tilde \Omega, \tilde \omega_{c})}} &\approx& \frac{1}{\mathcal B_1(\tilde \Omega, \tilde \omega_{c})} - 1, \quad \frac{e^{-\mathcal C_1(\tilde \Omega, \tilde \omega_{c})}}{1-e^{-\mathcal C_1(\tilde \Omega, \tilde \omega_{c})}} \approx \frac{1}{\mathcal C_1(\tilde \Omega, \tilde \omega_{c})} - 1.
		\end{eqnarray}
	}
	
	The critical derivative term becomes:
	\begin{eqnarray}
		\mathcal{D} &=& -\frac{\partial \mathcal A_1(\tilde \Omega)}{\partial B} + \frac{e^{-\mathcal B_1(\tilde \Omega, \tilde \omega_{c})}}{1-e^{-\mathcal B_1(\tilde \Omega, \tilde \omega_{c})}}\frac{\partial \mathcal B_1(\tilde \Omega, \tilde \omega_{c})}{\partial B} + \frac{e^{-\mathcal C_1(\tilde \Omega, \tilde \omega_{c})}}{1-e^{-\mathcal C_1(\tilde \Omega, \tilde \omega_{c})}}\frac{\partial \mathcal C_1(\tilde \Omega, \tilde \omega_{c})}{\partial B} \\ 
		&=& \left( \frac{1}{\mathcal B_1(\tilde \Omega, \tilde \omega_{c})}\frac{\partial \mathcal B_1(\tilde \Omega, \tilde \omega_{c})}{\partial B} + \frac{1}{\mathcal C_1(\tilde \Omega, \tilde \omega_{c})}\frac{\partial \mathcal C_1(\tilde \Omega, \tilde \omega_{c})}{\partial B} \right) - \left( \frac{\partial \mathcal A_1(\tilde \Omega)}{\partial B} + \frac{\partial \mathcal B_1(\tilde \Omega, \tilde \omega_{c})}{\partial B}\nonumber \right.\\ 
		&&\left.+ \frac{\partial \mathcal C_1(\tilde \Omega, \tilde \omega_{c})}{\partial B} \right).
	\end{eqnarray}
	
	Since \( \frac{\partial \mathcal B_1(\tilde \Omega, \tilde \omega_{c})}{\partial B} \propto \beta \) and \( \mathcal B_1(\tilde \Omega, \tilde \omega_{c}) \propto \beta \), the dominant contribution leads to
	\begin{equation}
		\mathcal{D} \approx \frac{1}{\mathcal B_1(\tilde \Omega, \tilde \omega_{c})}\frac{\partial \mathcal B_1(\tilde \Omega, \tilde \omega_{c})}{\partial B} + \frac{1}{\mathcal C_1(\tilde \Omega, \tilde \omega_{c})}\frac{\partial \mathcal C_1(\tilde \Omega, \tilde \omega_{c})}{\partial B} = \mathcal{C}_1,
	\end{equation}
	
	where \( \mathcal{C}_1 \) is temperature-independent.
	The overall temperature dependence is provided by
	\begin{equation}
		M_{1}^{(\theta)}(T \to \infty) \propto T^{1/2} \times T^{-2} = T^{-3/2}.
	\end{equation}
	
	Thereby,
	{\small \begin{equation}
			M_{1}^{(\theta)}(T \to \infty) \approx \frac{\mathcal{K}_H}{T^{3/2}},
		\end{equation}
	}
	
	where the constant \( \mathcal{K}_H \) is delivered as
	\begin{equation}
		\mathcal{K}_H = n \left(\frac{MV\omega_c}{2\pi\hbar}\right)^{-1} \left(\frac{M}{2\pi\hbar^2}\right)^{-1/2} k_B^{3/2} \ (\hbar \tilde{\Omega})^2 \left(1 - \frac{\tilde{\omega_c}^2}{\tilde{\Omega}^2}\right)\left( \frac{e}{Mc\omega_c} + \mathcal{C}_1 \right) e^{\mathcal A_1(\tilde \Omega)}.
	\end{equation}
	
	The magnetic susceptibility is given by
	\begin{eqnarray}
		\chi_{1}^{(\theta)} &=& -\frac{\partial M_{1}^{(\theta)}}{\partial B} =  \frac{n}{\beta} \left(\frac{MV\omega_c}{2\pi\hbar}\right)^{-1}\left(\frac{M}{2\pi\hbar^2\beta}\right)^{-1/2} \cdot (\mathfrak G^1(x, \tilde \Omega, \tilde \omega_{c}) + \mathfrak G^2(x, \tilde \Omega, \tilde \omega_{c})), \nonumber \\
	\end{eqnarray}
	
	where
	{\small\begin{eqnarray}
			\mathfrak G^1(x, \tilde \Omega, \tilde \omega_{c})& =& -\frac{\partial}{\partial B}\left(\frac{[1-e^{-\mathcal B_1(\tilde \Omega, \tilde \omega_{c})}][1-e^{-\mathcal C_1(\tilde \Omega, \tilde \omega_{c})}]}{e^{-\mathcal A_1(\tilde \Omega)}}\right) \cdot \mathfrak G^3(x, \tilde \Omega, \tilde \omega_{c}), \cr
			\mathfrak G^2(x, \tilde \Omega, \tilde \omega_{c}) &=& \frac{[1-e^{-\mathcal B_1(\tilde \Omega, \tilde \omega_{c})}][1-e^{-\beta_2}]}{e^{-\mathcal A_1(\tilde \Omega)}} \cdot \frac{\partial \mathfrak G^3(x, \tilde \Omega, \tilde \omega_{c})}{\partial B}, \cr
			\mathfrak G^3(x, \tilde \Omega, \tilde \omega_{c})& =& \frac{e}{Mc\omega_c} - \frac{\partial \mathcal A_1(\tilde \Omega)}{\partial B} + \frac{e^{-\mathcal B_1(\tilde \Omega, \tilde \omega_{c})}}{1-e^{-\mathcal B_1(\tilde \Omega, \tilde \omega_{c})}}\frac{\partial \mathcal B_1(\tilde \Omega, \tilde \omega_{c})}{\partial B}\cr 
			&&+ \frac{e^{-\mathcal C_1(\tilde \Omega, \tilde \omega_{c})}}{1-e^{-\mathcal C_1(\tilde \Omega, \tilde \omega_{c})}}\frac{\partial \mathcal C_1(\tilde \Omega, \tilde \omega_{c})}{\partial B}.
		\end{eqnarray}
	}

	
	In addition, we have the following expansions
	{\small \begin{eqnarray}
			\frac{e^{-\mathcal B_1(\tilde \Omega, \tilde \omega_{c})}}{1-e^{-\mathcal B_1(\tilde \Omega, \tilde \omega_{c})}} &=& \frac{1 - \mathcal B_1(\tilde \Omega, \tilde \omega_{c}) + \frac{\mathcal B_1(\tilde \Omega, \tilde \omega_{c})^2}{2} + O(\mathcal B_1(\tilde \Omega, \tilde \omega_{c})^3)}{\mathcal B_1(\tilde \Omega, \tilde \omega_{c}) - \frac{\mathcal B_1(\tilde \Omega, \tilde \omega_{c})^2}{2} + \frac{\mathcal B_1(\tilde \Omega, \tilde \omega_{c})^3}{6} + O(\mathcal B_1(\tilde \Omega, \tilde \omega_{c})^4)}\cr 
			&=& \frac{1}{\mathcal B_1(\tilde \Omega, \tilde \omega_{c})} \left[1 - \frac{\mathcal B_1(\tilde \Omega, \tilde \omega_{c})}{2} + \frac{\mathcal B_1(\tilde \Omega, \tilde \omega_{c})^2}{12} + O(\mathcal B_1(\tilde \Omega, \tilde \omega_{c})^3)\right].
		\end{eqnarray}
	}
	
	Similarly, we have
	\begin{equation}
		\frac{e^{-\mathcal C_1(\tilde \Omega, \tilde \omega_{c})}}{1-e^{-\mathcal C_1(\tilde \Omega, \tilde \omega_{c})}} = \frac{1}{\mathcal C_1(\tilde \Omega, \tilde \omega_{c})} \left[1 - \frac{\mathcal C_1(\tilde \Omega, \tilde \omega_{c})}{2} + \frac{\mathcal C_1(\tilde \Omega, \tilde \omega_{c})^2}{12} + O(\mathcal C_1(\tilde \Omega, \tilde \omega_{c})^3)\right],
	\end{equation}
	{\small \begin{eqnarray}
			\chi_{1}^{(\theta)} &\propto&  T^{1/2} \cdot (T^{-2} + T^{-3}) \propto T^{-3/2} + T^{-5/2},
		\end{eqnarray}
	}
	
	where the dominant term leads to
	{\small \begin{equation}
			\chi_{1}^{(\theta)}(T \to \infty) \propto T^{-3/2}.
		\end{equation}
	}
	
	\item {\bf Case of low temperature limit ($T \to 0$, $x \to \infty$).}
	
	When $x \gg 1$, we obtain the asymptotic behavior by examining each component of the partition function.
	The exponential term in the numerator of (\ref{partidef000}) behaves as:
	{\small \begin{equation}
			e^{-x\left[\frac{1}{2}-\frac{e}{2\hbar\tilde{\Omega}}\left(E_{1}x_{0}+E_{2}y_{0}\right)\right]} \approx e^{-\frac{x}{2}} e^{\left[\frac{ex}{2\hbar\tilde{\Omega}}(E_{1}x_{0}+E_{2}y_{0})\right]},
		\end{equation}
	}
	
	while in the denominator, each factor approaches unity, since $1-e^{-\alpha x/2} \approx 1$ when $x \gg 1$, such that
	{\small\begin{equation}
			\left[1-e^{-\frac{x}{2}\left(1-\frac{\tilde{\omega}_{c}}{\tilde{\Omega}}\right)}\right]\left[1-e^{-\frac{x}{2}\left(1+\frac{\tilde{\omega}_{c}}{\tilde{\Omega}}\right)}\right] \approx 1.
		\end{equation}
	}
	
	The partition function (\ref{partidef000}), then, reduces to
	{\small \begin{equation}
			Z_{1}^{(\theta)} \approx V\left(\frac{M\omega_{c}}{2\pi\hbar}\right)\left(\frac{M}{2\pi\hbar^{2}\beta}\right)^{\frac{1}{2}} e^{-\frac{x}{2}} e^{\left[\frac{ex}{2\hbar\tilde{\Omega}}(E_{1}x_{0}+E_{2}y_{0})\right]}.
		\end{equation}
	}
	
	This result reflects the ground state dominance with zero-point energy $\frac{\hbar\tilde{\Omega}}{2}$ and an energy shift proportional to the external electric fields $(E_1, E_2)$ acting on the guiding center coordinates $(x_0, y_0)$.
	The first logarithmic term in the free energy (\ref{frreeenrg000}) rewrites
	{\small \begin{equation}
			-\frac{n}{\beta} \ln\left(\frac{MV\omega_{c}}{2\pi\hbar}\sqrt{\frac{M}{2\pi\hbar^{2}\beta}}\right) = -\frac{n}{\beta} \ln\left(\frac{MV\omega_{c}}{2\pi\hbar}\right) - \frac{n}{2\beta} \ln\left(\frac{M}{2\pi\hbar^{2}\beta}\right).
		\end{equation}
	}
	
	As $\beta \to \infty$, this term vanishes as $O(\ln\beta/\beta)$.
	
	For the second logarithmic term in (\ref{frreeenrg000}), since $\beta\hbar(\tilde{\Omega} \pm \tilde{\omega}_c) \gg 1$, it comes
	{\small \begin{equation}
			\ln\left[ \left(1-e^{-\beta\hbar(\tilde{\Omega}-\tilde{\omega}_c)}\right)\left(1-e^{-\beta\hbar(\tilde{\Omega}+\tilde{\omega}_c)}\right) \right] \approx 0.
		\end{equation}
	}
	
	Thereby, the dominant terms in the low temperature limit for the free energy (\ref{frreeenrg000}) lead to
	{\small \begin{equation}
			F_{1}^{(\theta)} \approx \frac{n\hbar\tilde{\Omega}}{2} - \frac{ne(E_{1}x_{0}+E_{2}y_{0})}{2}.
		\end{equation}
	}
	
\end{itemize}

The derivative term simplifies as follows:
{\small \begin{equation}
		\mathcal{D} = -\frac{\partial \mathcal A_1(\tilde \Omega)}{\partial B} + \underbrace{\frac{e^{-\mathcal B_1(\tilde \Omega, \tilde \omega_{c})}}{1-e^{-\mathcal B_1(\tilde \Omega, \tilde \omega_{c})}}\frac{\partial \mathcal B_1(\tilde \Omega, \tilde \omega_{c})}{\partial B}}_{\approx 0} + \underbrace{\frac{e^{-\mathcal C_1(\tilde \Omega, \tilde \omega_{c})}}{1-e^{-\mathcal C_1(\tilde \Omega, \tilde \omega_{c})}}\frac{\partial \mathcal C_1(\tilde \Omega, \tilde \omega_{c})}{\partial B}}_{\approx 0} \approx -\frac{\partial \mathcal A_1(\tilde \Omega)}{\partial B}.
	\end{equation}
}

The leading temperature dependence is supplied by
{\small \begin{equation}
		M_{1}^{(\theta)}(T \to 0) \approx \mathcal{K}_L \cdot T^{1/2},
	\end{equation}
}

where the constant \( \mathcal{K}_L \) is such that
{\small 
	\begin{eqnarray}
		\mathcal{K}_L &=& n \left(\frac{MV\omega_c}{2\pi\hbar}\right)^{-1} \left(\frac{M}{2\pi\hbar^2}\right)^{-1/2} k_B^{1/2} \ e^{\mathcal A_1(\tilde \Omega)}\left(\frac{e}{Mc\omega_c} - \frac{\partial \mathcal A_1(\tilde \Omega)}{\partial B}\right).
	\end{eqnarray}
}

For $\beta\hbar\tilde{\Omega} \gg 1$, we have the following approximations
{\small 
	\begin{align}
		e^{-\mathcal B_1(\tilde \Omega, \tilde \omega_{c})} &\ll 1, \quad e^{-\mathcal C_1(\tilde \Omega, \tilde \omega_{c})} \ll 1,\quad 1 - e^{-\mathcal B_1(\tilde \Omega, \tilde \omega_{c})} \approx 1, \quad 1 - e^{-\mathcal C_1(\tilde \Omega, \tilde \omega_{c})} \approx 1, \nonumber \\ \frac{e^{-\mathcal B_1(\tilde \Omega, \tilde \omega_{c})}}{1-e^{-\mathcal B_1(\tilde \Omega, \tilde \omega_{c})}} &\approx e^{-\mathcal B_1(\tilde \Omega, \tilde \omega_{c})} \ll 1, \quad \frac{e^{-\mathcal C_1(\tilde \Omega, \tilde \omega_{c})}}{1-e^{-\mathcal C_1(\tilde \Omega, \tilde \omega_{c})}} \approx e^{-\mathcal C_1(\tilde \Omega, \tilde \omega_{c})} \ll 1.
	\end{align}
}

Besides, we have
{\small 
	\begin{align}
		\chi_{1}^{(\theta)} &= \propto T^{1/2} \cdot (T^{-1} + T^{-1})\quad \propto T^{-1/2}.
	\end{align}
}

\section{Concluding remarks}

This work presents a comprehensive theoretical framework for analyzing the thermodynamic properties of a two-dimensional electron gas in non-commutative spaces, within the context of Tsallis non-extensive statistics. The study establishes several fundamental contributions at the intersection of quantum mechanics, statistical physics, and non-commutative geometry. The central achievement of this research lies in the rigorous derivation of analytical expressions for key thermodynamic quantities, including the partition function, free energy, magnetization, magnetic susceptibility, and specific heat, which explicitly depend on both the non-extensivity parameter $q$ and the non-commutativity parameter $\theta$. Through the systematic adaptation of the Hilhorst integral transform to non-commutative geometry, we have demonstrated that the combined effects of spatial non-commutativity and statistical non-extensivity give rise to fundamentally new thermodynamic regimes that cannot be captured by conventional approaches. Our analysis reveals that the non-commutativity parameter $\theta$ induces structural and non-perturbative modifications to the system's thermodynamic behavior. Remarkably, even infinitesimal deviations from the commutative limit ($\theta \to 0$) lead to drastic changes in all thermodynamic properties, indicating that non-commutativity represents a fundamental alteration of spacetime structure rather than a mere perturbative correction \cite{Connes1994}. The singular nature of the $q \to 1$ limit in non-commutative space further underscores that standard Boltzmann-Gibbs statistics cannot be continuously recovered from the non-extensive and non-commutative framework \cite{Ubriaco2016}. Temperature-dependent analysis demonstrates distinct physical regimes: at low temperatures, non-commutativity modifies ground state energies and quantum degeneracies, while at high temperatures, the effective phase space volume and classical scaling laws ($T^{5/2}$) are altered by non-commutative corrections. The interplay between $\theta$ and external electromagnetic fields produces anomalous responses in magnetization and susceptibility, potentially signaling new phase transitions or emergent collective behaviors. From a broader perspective, this work establishes that the combination of non-commutative geometry, characterized by a fundamental minimal length, and non-extensive statistical mechanics, incorporating long-range interactions and non-additive entropy, fundamentally reshapes the thermody
namics of quantum electronic systems. These effects are not mere quantitative modifications but represent qualitative departures from standard theory, opening new theoretical and experimental directions in condensed matter physics, quantum information theory, and fundamental physics research \cite{Rivasseau2007}. The mathematical framework developed here, particularly the adapted Hilhorst transform, provides a powerful analytical bridge between non-extensive and extensive ensembles in non-commutative geometry. This methodology can be extended to other quantum systems where both spatial discreteness and non-standard statistical behaviors play essential roles, potentially revealing new physics in systems ranging from quantum dots and graphene to cosmological models and quantum gravity scenarios \cite{Konishi2020}. Future investigations may explore experimental signatures of these non-commutative
tive and non-extensive effects in real condensed matter systems, as well as the extension of this framework to higher dimensions and more complex field configurations \cite{Calmet2021}. The singular relationship between commutativity and extensivity highlighted in this work suggests that non-commutative quantum systems might intrinsically require a non-extensive statistical treatment, a principle with profound implications for our understanding of quantum thermodynamics at fundamental scales \cite{Luciano2022}.
 
 \medskip
 \textbf{Acknowledgements} \par 
 The authors gratefully acknowledge CERME (Centre d\textquoteright Excellence Régional pour la 
 
 Maîtrise de  l\textquoteright Electricité) for providing the necessary computer equipment that made 
 
 this work possible.

\end{document}